\documentclass[useAMS,usenatbib]{mn2e}

\usepackage{amsmath,amssymb,graphicx}

\newcommand{\apjs}{ApJS}
\newcommand{\mnras}{MNRAS}
\newcommand{\apj}{ApJ}
\newcommand{\aap}{A\&A}

\newcommand{\icarus}{Icarus}

\newcommand{\kepler}{\textit{Kepler}}

\newcommand{\xxii}{$\Xi$}
\newcommand{\ximax}{$\Xi_\text{max}$}

\newcommand{\koiofe}{KOI-148}
\newcommand{\koitfe}{KOI-248}
\newcommand{\koitst}{KOI-262}
\newcommand{\koistz}{KOI-620}
\newcommand{\koissf}{KOI-775}
\newcommand{\koietn}{KOI-829}
\newcommand{\koiees}{KOI-886}
\newcommand{\koinzf}{KOI-904}
\newcommand{\koiotfo}{KOI-1241}
\newcommand{\koiotsz}{KOI-1270}
\newcommand{\koiotts}{KOI-1336}
\newcommand{\koioftn}{KOI-1529}
\newcommand{\koitzes}{KOI-2086}
\newcommand{\kepofe}{Kepler-48}
\newcommand{\keptfe}{Kepler-49}
\newcommand{\keptst}{Kepler-50}
\newcommand{\kepstz}{Kepler-51}
\newcommand{\kepssf}{Kepler-52}
\newcommand{\kepetn}{Kepler-53}
\newcommand{\kepees}{Kepler-54}
\newcommand{\kepnzf}{Kepler-55}
\newcommand{\kepotfo}{Kepler-56}
\newcommand{\kepotsz}{Kepler-57}
\newcommand{\kepotts}{Kepler-58}
\newcommand{\kepoftn}{Kepler-59}
\newcommand{\keptzes}{Kepler-60}

\newcommand{\nsystems}{13}
\newcommand{\nconfirm}{27}

\title[Kepler planetary systems 48 to 60]{Transit Timing Observations from \kepler: VII.  Confirmation of \nconfirm\ planets in \nsystems\ multiplanet systems via Transit Timing Variations and orbital stability}
\author[Steffen \textit{et al.}]{
Jason H. Steffen$^{1}$,
Daniel C. Fabrycky$^{2,3}$,
Eric Agol$^{4}$,
Eric B. Ford$^{5}$,\newauthor
Robert C. Morehead$^{5,6}$,
William D. Cochran$^{7}$,
Jack J. Lissauer$^{8}$,\newauthor
Elisabeth R. Adams$^{9}$,
William J. Borucki$^{8}$,
Steve Bryson$^{8}$,
Douglas A. Caldwell$^{10}$,\newauthor
Andrea Dupree$^{9}$,
Jon M. Jenkins$^{8}$, 
Paul Robertson$^{7}$,
Jason F. Rowe$^{8}$,\newauthor
Shawn Seader$^{8}$,
Susan Thompson$^{8}$, and
Joseph D. Twicken$^{8}$
\\
$^{1}$Fermilab Center for Particle Astrophysics, P.O. Box 500, MS 127, Batavia, IL 60510\\
$^{2}$UCO/Lick Observatory, University of California, Santa Cruz, CA 95064, USA\\
$^{3}$Hubble Fellow\\
$^{4}$Department of Astronomy, University of Washington, Seattle, WA 90195\\
$^{5}$Astronomy Department, University of Florida, 211 Bryant Space Sciences Center, Gainesville, FL 32111, USA\\
$^{6}$National  Science Foundation Graduate Research Fellow\\
$^{7}$McDonald Observatory, The University of Texas, Austin TX 78730, USA\\
$^{8}$NASA Ames Research Center, Moffett Field, CA, 94035, USA\\
$^{9}$Harvard-Smithsonian Center for Astrophysics, 60 Garden Street, Cambridge, MA 02138, USA\\
$^{10}$SETI Institute, Mountain View, CA, 94043, USA\\
}

\begin{document}

%\date{Accepted 1988 December 15. Received 1988 December 14; in original form 1988 October 11}

\pagerange{\pageref{firstpage}--\pageref{lastpage}} %\pubyear{2002}

\maketitle

\label{firstpage}

\begin{abstract}
We confirm \nconfirm\ planets in \nsystems\ planetary systems by showing the existence of statistically significant anti-correlated transit timing variations (TTVs), which demonstrates that the planet candidates are in the same system, and long-term dynamical stability, which places limits on the masses of the candidates---showing that they are planetary.  %This overall method of planet confirmation was first applied to \kepler\ systems 23 through 32.  
All of these newly confirmed planetary systems have orbital periods that place them near first-order mean motion resonances (MMRs), including 6 systems near the 2:1 MMR, 5 near 3:2, and one each near 4:3, 5:4, and 6:5.  In addition, several unconfirmed planet candidates exist in some systems (that cannot be confirmed with this method at this time).  A few of these candidates would also be near first order MMRs with either the confirmed planets or with other candidates.  One system of particular interest, \kepotfo\ (\koiotfo ), is a pair of planets orbiting a 12th magnitude, giant star with radius over three times that of the Sun and effective temperature of 4900~K---among the largest stars known to host a transiting exoplanetary system.
\end{abstract}

\begin{keywords}
celestial mechanics; methods: data analysis; techniques: photometric
\end{keywords}

\section{Introduction}

NASA's \kepler\ mission continues to identify many candidate transiting exoplanet systems, which now number nearly 2000 \citep{Borucki:2011,Batalha:2012}.  The process of confirming or validating these planet candidates as real exoplanets often requires a significant amount of analysis and effort; and consequently the number of confirmed planets lags far behind the new candidate discoveries.  In an effort to ameliorate this situation we published a new approach to planet confirmation that requires a less detailed analysis \citep{Fabrycky:2012a,Ford:2012a,Steffen:2012a} and can generally be accomplished in much less time.  This approach essentially relies on demonstrating that two transiting candidates are in the same system and that their masses are planetary.  Both aspects of this confirmation method rely on dynamical interactions.  First, planets are shown to be in the same system by looking for anticorrelated transit timing variations (TTVs) that arise from short-term changes in obital period due to planet-planet interactions within the system \citep{Agol:2005,Holman:2005}.  Second, the candidates are shown to have planetary mass by requiring that the system be dynamically stable.

This method is a particularly useful means of identifying true planetary systems from among the false positive systems---allowing ground-based follow-up resources to be devoted to studying systems that are both real and dynamically interesting (as manifest by their significant TTV signal).  While detailed TTV analyses have been instrumental in confirming several multiplanet systems (e.g., Kepler 9 \citep{Holman:2010}, Kepler 11 \citep{Lissauer:2011a}, Kepler 18 \citep{Cochran:2011}, and Kepler 36 \citep{Carter:2012}), 21 planets in 10 multiplanet systems have been confirmed by these new methods that we apply again here (Kepler systems 23 through 32, \citet{Fabrycky:2012a,Ford:2012a,Steffen:2012a}).  These 10 confirmed multiplanet systems include several that have particularly interesting properties and are likely to be the subject of future investigations.  For example: all of the systems are near mean-motion resonance (MMR), Kepler-25 (KOI-244, \citet{Steffen:2012a}) is relatively bright at Kp $\sim 10$, Kepler-30 (KOI-806, \citet{Fabrycky:2012a}) has TTVs that deviate from a constant period by nearly a day over the course of the \kepler\ data, and many (Kepler 23, 24, 26, 31, and 32) have additional planet candidates that may be confirmed or validated.  As more data from \kepler\ are gathered, new systems that can be confirmed by the means outlined above can be found.  In this paper we apply the same methods used in \citep{Fabrycky:2012a,Ford:2012a,Steffen:2012a} to 2 additional quarters of \kepler\ data (all data through Q8) and confirm \nconfirm\ new planets in \nsystems\ systems.

This paper is organized as follows.  In \S \ref{secProperties} we summarize the stellar properties for the host stars in these systems and the basic orbital and physical characteristics of the planets.  The dynamical confirmation of the relevant \nconfirm\ planets is presented in \S \ref{secConfirm}.  % and the validation of the remaining \nvalidate\ planets is in \S \ref{secValidate}.  
We make concluding remarks in \S \ref{secConclusion}.  Data and results from some of the analyses performed in this paper are shown in the Appendix.

\section{Stellar and Planet Properties}\label{secProperties}

We take the stellar properties from the tables given in \citet{Batalha:2012}.  The procedure used for determining the stellar parameters (effective temperature Teff, surface gravity $\log g$ and stellar radius $R_\star$) for our systems was discussed in detail in Section 5.2 of that paper. In most cases, Teff and $\log g$ that were determined photometrically and reported in the \kepler\ Input Catalog were used as a starting point.  These values were then corrected by matching to the Yonsei-Yale stellar evolution models \citep{Demarque:2004} (cf. Figure 2 of \citet{Batalha:2012}).  \kepler\ Followup Observing Program reconnaissance spectra have also been obtained for most of these KOIs.  In general there is reasonable agreement between the stellar parameters in Table 1 and those obtained through the reconnaissance spectroscopy.  In a few cases of disagreement, the discrepancy can be attributed to the low S/N of the reconnaissance spectra, thus we rely on the published parameters.  The basic planet size and orbital properties for the systems under consideration are also taken from \citep{Batalha:2012} and are given in Table \ref{tabPlanets}\footnote{We note that because these planets show significant TTVs, the estimates for the transit depth and duration will be off slightly since the \kepler\ pipeline assumes a linear ephemeris.}.  We note that the giant star \koiotfo\ along with \koitst\ are quite bright and therefore amenable to in-depth asteroseismic study.  We expect that the stellar properties, particularly for the giant, may be revised somewhat from the values presented here and in \citet{Batalha:2012} once such studies have been completed.  In neither case is the planet interpretation of the transiting candidates in jeopardy.

\begin{table}
\centering
\caption{Table of stellar properties for the planetary systems (taken from \citet{Batalha:2012}).}
\begin{tabular}{lllccccc}
\hline
Kepler	&	KOI	&	KID	&	Kp	&	Teff	&	$\log g$	&	$R_\star$	&	$M_\star$	\\
	&		&		&		&	(K)	&		&	($R_\odot$)	&	($M_\odot$)	\\	\hline \hline
48	&	148	&	5735762	&	13.04	&	5190	&	4.49	&	0.89	&	0.89	\\	
49	&	248	&	5364071	&	15.264	&	3974	&	4.73	&	0.53	&	0.55	\\	
50	&	262	&	11807274	&	10.421	&	6058	&	3.98	&	1.88	&	1.23	\\	
51	&	620	&	11773022	&	14.669	&	5803	&	4.52	&	0.91	&	1.00	\\	
52	&	775	&	11754553	&	15.095	&	4075	&	4.74	&	0.52	&	0.54	\\	
53	&	829	&	5358241	&	15.386	&	5858	&	4.53	&	0.89	&	0.98	\\	
54	&	886	&	7455287	&	15.847	&	3705	&	4.75	&	0.5	&	0.51	\\	
55	&	904	&	8150320	&	15.791	&	4362	&	4.7	&	0.58	&	0.62	\\	
56 	&	1241	&	6448890	&	12.44	&	4931	&	3.58	&	3.14	&	1.37	\\
57	&	1270	&	8564587	&	14.809	&	5145	&	4.63	&	0.73	&	0.83	\\	
58	&	1336	&	4077526	&	14.82	&	5843	&	4.39	&	1.03	&	0.95	\\	
59	&	1529	&	9821454	&	14.307	&	6074	&	4.51	&	0.94	&	1.04	\\	
60	&	2086	&	6768394	&	13.959	&	5915	&	4.13	&	1.5	&	1.11	\\	\hline
\end{tabular}
\label{tabStars}
\end{table}

\begin{table*}
\centering
\begin{minipage}{180mm}
\caption{Table of planet orbital properties for the planetary systems.}
\begin{tabular}{llccccccccccc}
\hline
Planet		&	KOI	&	$r/R_\star$	&	$\sigma_{r/R_\star}$	&	Radius	&	Duration	&	$T_0$	&	$\sigma_{T_0}$	&	period	&	$\sigma_{\text{period}}$	&	$P_i/P_1$	&	$P_i/P_{i-1}$	&	$\xi$	\\	
		&		&		&		&	($R_\oplus$)	&	(hours)	&	(days)	&	(days)	&	(days)	&	(days)	&		&		&		\\	\hline \hline
48	b	&	148.01	&	0.02208	&	3.1E-04	&	2.14	&	2.8412	&	57.06301	&	1.0E-03	&	4.7779803	&	1.4E-05	&	1.0000	&		&		\\	
48	c	&	148.02	&	0.03243	&	4.0E-04	&	3.14	&	3.5027	&	58.33981	&	7.7E-04	&	9.6739283	&	2.0E-05	&	2.0247	&	2.0247	&	1.03	\\	
		&	148.03	&	0.02426	&	4.1E-04	&	2.35	&	5.9308	&	79.06567	&	2.8E-03	&	42.895916	&	3.2E-04	&	8.9778	&	4.4342	&	0.97	\\	\hline
		&	248.03	&	0.03213	&	1.4E-03	&	1.85	&	1.8736	&	105.12838	&	1.2E-03	&	2.576555	&	1.2E-05	&	1.0000	&		&		\\	
49	b	&	248.01	&	0.04723	&	1.9E-03	&	2.72	&	2.9766	&	103.28753	&	1.0E-03	&	7.2037945	&	2.2E-05	&	2.7959	&	2.7959	&	0.89	\\	
49	c	&	248.02	&	0.04423	&	2.0E-03	&	2.55	&	2.8577	&	102.84919	&	2.2E-03	&	10.9129343	&	6.9E-05	&	4.2355	&	1.5149	&	1.20	\\	
		&	248.04	&	0.03445	&	2.0E-03	&	1.99	&	2.8905	&	74.26167	&	4.3E-03	&	18.5962181	&	2.1E-04	&	7.2175	&	1.7041	&	1.18	\\	\hline
50	b	&	262.01	&	0.01074	&	1.5E-04	&	2.2	&	4.2439	&	105.63186	&	2.2E-03	&	7.8125124	&	5.2E-05	&	1.0000	&		&		\\	
50	c	&	262.02	&	0.01362	&	3.0E-04	&	2.79	&	2.6022	&	69.96857	&	2.2E-03	&	9.3761368	&	5.6E-05	&	1.2001	&	1.2001	&	1.73	\\	\hline
51	b	&	620.01	&	0.07074	&	2.0E-04	&	7.05	&	5.7799	&	92.10435	&	6.2E-04	&	45.1555023	&	7.2E-05	&	1.0000	&		&		\\	
51	c	&	620.03	&	0.05731	&	8.1E-03	&	5.71	&	2.7849	&	228.32095	&	2.2E-03	&	85.3128662	&	9.6E-04	&	1.8893	&	1.8893	&	2.57	\\	
		&	620.02	&	0.09717	&	2.4E-04	&	9.68	&	8.4192	&	145.02345	&	6.2E-04	&	130.1831055	&	3.3E-04	&	2.8830	&	1.5259	&	0.38	\\	\hline
52	b	&	775.02	&	0.03698	&	3.8E-03	&	2.1	&	2.4223	&	109.37692	&	1.5E-03	&	7.8773565	&	3.8E-05	&	1.0000	&		&		\\	
52	c	&	775.01	&	0.03225	&	3.3E-03	&	1.84	&	3.1252	&	105.72549	&	2.7E-03	&	16.3850021	&	1.4E-04	&	2.0800	&	2.0800	&	0.99	\\	
		&	775.03	&	0.03268	&	3.4E-03	&	1.86	&	4.1756	&	84.59631	&	4.4E-03	&	36.4459076	&	4.9E-04	&	4.6267	&	2.2243	&	0.98	\\	\hline
		&	829.02	&	0.01987	&	8.4E-04	&	1.92	&	4.0291	&	71.78461	&	4.8E-03	&	9.7519283	&	1.3E-04	&	1.0000	&		&		\\	
53	b	&	829.01	&	0.02987	&	1.1E-03	&	2.89	&	4.4815	&	107.77608	&	2.9E-03	&	18.6489525	&	1.6E-04	&	1.9123	&	1.9123	&	1.12	\\	
53	c	&	829.03	&	0.03277	&	1.2E-03	&	3.17	&	5.0162	&	96.84816	&	3.8E-03	&	38.5583038	&	4.1E-04	&	3.9539	&	2.0676	&	1.14	\\	\hline
54	b	&	886.01	&	0.03842	&	1.3E-03	&	2.1	&	3.4129	&	103.20573	&	2.9E-03	&	8.0109434	&	6.9E-05	&	1.0000	&		&		\\	
54	c	&	886.02	&	0.02254	&	1.0E-03	&	1.23	&	4.9738	&	76.44842	&	9.1E-03	&	12.0717249	&	2.9E-04	&	1.5069	&	1.5069	&	0.79	\\	
		&	886.03	&	0.0318	&	4.7E-03	&	1.74	&	3.8411	&	85.32592	&	9.7E-03	&	20.9954014	&	5.8E-04	&	2.6208	&	1.7392	&	1.56	\\	\hline
		&	904.01	&	0.02607	&	2.7E-03	&	1.65	&	1.6829	&	103.14939	&	1.4E-03	&	2.2111206	&	1.2E-05	&	1.0000	&		&		\\	
		&	904.04	&	0.02306	&	2.5E-03	&	1.46	&	2.1085	&	68.04531	&	3.1E-03	&	4.6175265	&	4.2E-05	&	2.0883	&	2.0883	&	1.02	\\	
		&	904.05	&	0.02689	&	2.8E-03	&	1.7	&	2.3911	&	70.91412	&	4.2E-03	&	10.1986198	&	1.1E-04	&	4.6124	&	2.2087	&	1.15	\\	
55	b	&	904.02	&	0.03832	&	3.9E-03	&	2.43	&	3.4932	&	111.79598	&	3.1E-03	&	27.9481449	&	2.4E-04	&	12.6398	&	2.7404	&	0.96	\\	
55	c	&	904.03	&	0.0349	&	3.7E-03	&	2.21	&	5.0617	&	71.99284	&	4.5E-03	&	42.1516418	&	5.3E-04	&	19.0635	&	1.5082	&	0.79	\\	\hline
56	b	&	1241.02	&	0.0112	&	3.3E-04	&	3.84	&	13.5401	&	68.55893	&	8.8E-03	&	10.5034294	&	2.5E-04	&	1.0000	&		&		\\	
56	c	&	1241.01	&	0.02292	&	3.3E-04	&	7.85	&	11.0797	&	80.0928	&	6.1E-03	&	21.4050484	&	3.6E-04	&	2.0379	&	2.0379	&	1.55	\\	\hline
57	b	&	1270.01	&	0.02762	&	6.6E-03	&	2.19	&	1.1185	&	71.5584	&	1.0E-03	&	5.7293196	&	1.8E-05	&	1.0000	&		&		\\	
57	c	&	1270.02	&	0.01958	&	4.6E-03	&	1.55	&	1.7074	&	66.99083	&	2.7E-03	&	11.6092567	&	8.6E-05	&	2.0263	&	2.0263	&	0.83	\\	\hline
58	b	&	1336.01	&	0.02471	&	1.0E-03	&	2.78	&	4.3967	&	69.87048	&	5.6E-03	&	10.2184954	&	1.5E-04	&	1.0000	&		&		\\	
58	c	&	1336.02	&	0.02542	&	1.1E-03	&	2.86	&	5.1013	&	65.43209	&	6.6E-03	&	15.5741568	&	2.7E-04	&	1.5241	&	1.5241	&	0.99	\\	\hline
59	b	&	1529.02	&	0.01065	&	5.3E-04	&	1.1	&	5.977	&	70.74767	&	1.1E-02	&	11.8681707	&	3.6E-04	&	1.0000	&		&		\\	
59	c	&	1529.01	&	0.01923	&	7.6E-04	&	1.98	&	4.1249	&	80.69818	&	6.5E-03	&	17.9801235	&	3.0E-04	&	1.5150	&	1.5150	&	1.66	\\	\hline
60	b	&	2086.01	&	0.01397	&	3.9E-04	&	2.28	&	4.1519	&	69.03637	&	4.7E-03	&	7.1316185	&	9.3E-05	&	1.0000	&		&		\\	
60	c	&	2086.02	&	0.01513	&	4.0E-04	&	2.47	&	4.4059	&	73.06513	&	4.7E-03	&	8.9193459	&	1.1E-04	&	1.2507	&	1.2507	&	1.02	\\	
60	d	&	2086.03	&	0.0156	&	8.7E-04	&	2.55	&	2.9856	&	66.24626	&	6.3E-03	&	11.9016171	&	2.0E-04	&	1.6689	&	1.3344	&	1.62	\\	\hline
\end{tabular}
\label{tabPlanets}
\end{minipage}
\end{table*}

\section{Confirmation via TTVs and orbital stability}\label{secConfirm}

\subsection{Anticorrelated TTVs}

As mentioned above and discussed in some detail in \citep{Fabrycky:2012a,Ford:2012a,Steffen:2012a}, the presence of anticorrelated TTV signatures among planet candidates on a single target is a strong indicator that the objects are in the same system.  Multiple methods can be used to establish these correlations---the approaches that we have used in the past include using Gaussian Process (GP) modeling \citep{Ford:2012a}, Fourier analysis methods \citep{Steffen:2012a}, and the use of simple dynamical modeling of the potential planetary system \citep{Fabrycky:2012a}.  Each of these approaches relies on differing amounts of physical underpinning (dynamical modeling applying the most physics while GP the least).  They also use different fundamental assumptions regarding the specific form of the TTV signal (GP being the most general of the three while dynamical modeling is the least).  The Fourier method lies between the other two methods both in terms of the assumed properties of the TTV signal and its physical underpinnings.  Consequently, these different methods are quite complementary---all are suitable for confirming some systems but none are suitable for confirming all systems that lie within the reach of this overall approach.

The GP method of \citet{Ford:2012a} models the TTV signal as a Gaussian process, deriving a generalized functional form for each object in a given pair.  From these GPs the correlation between TTV signals of the two planet candidates is calculated.  Finally, we estimate the probability that the observed anticorrelation is due to timing noise with a Monte Carlo analysis.  For this analysis we scramble the TTV residuals, remove any resulting linear trend, model the new residuals as a GP, and recalculate the correlation coefficient.  We require, as a threshold to claim that the correlation is dynamical in nature, that less than 0.1\% of the Monte Carlo realizations have a more significant correlation than the actual data.

The Fourier method of \citet{Steffen:2012a} generates a periodogram of the TTV signal by fitting sinusoidal models on a grid of periods:
\begin{equation}
f_i = A \sin \left(\frac{2\pi t}{P_i} \right) + B \cos \left(\frac{2\pi t}{P_i} \right) + C.
\label{fourcomp}
\end{equation}
where $A$, $B$, and $C$ are model parameters, and $P_i$ is the test timescale\footnote{This formula, originally given in \citet{Steffen:2012a}, had a typo that has been corrected here and in \citet{Steffen:2012c}. Specifically, $P/P_i$ is replaced by $1/P_i$ in the argument of the sinusoid functions.}.  The fitted values for $A$ and $B$ are stored along with their measured uncertainties, $\sigma_A$ and $\sigma_B$, derived from the covariance matrix of the three-parameter fit.  Next, a quantity \xxii\
\begin{equation}
\Xi = -\left(\frac{A_1A_2}{\sigma_{A1}\sigma_{A2}} + \frac{B_1B_2}{\sigma_{B1}\sigma_{B2}}\right)
\label{xistat}
\end{equation}
is calculated for each of the sampled periods where the ``1'' and ``2'' subscripts correspond to the two objects.  Once these values have been calculated, the maximum value of \xxii\ for a given candidate pair, \ximax, is recorded.  Finally, a Monte Carlo analysis is done to identify the probability of randomly finding a \ximax\ as large as that observed.  Again, our threshold for claiming evidence for dynamical interactions is 0.1\%.

The dynamical modeling of \citet{Fabrycky:2012a} seeks to extract the dominant TTV signal from numerical integration, and determine if this signal is present in the data.  A numerical integration of the system is run for each pair of planets in Table \ref{tabPlanets}.  We assume circular, coplanar orbits with the same orbital properties (period, phase) and physical properties (mass, derived from radius via the power-law $M_p = M_\oplus (R_p/R_\oplus)^{2.06}$ as was used in \citet{Lissauer:2011b}) as the planets we are modelling.  From that simulation, the dominant frequency of the resulting TTV signal is found in a periodogram \citep{Lomb:1976}.  We fit a sinusoid of that frequency to the data, determining its amplitude.  Again, a Monte Carlo study is done to see if a random realization can produce that large an amplitude at the theoretically-predicted frequency.  One unique characteristic of the dynamical modeling, contrasted with the GP and Fourier methods, is that it does not require a significant TTV signal from both objects in the tested pair---and therefore not requiring an anti-correlation in the timing residuals.  Rather, we only use the transits of the perturbing planet to determine its period and phase, from which the perturbed planet's expected TTV signal follows.  Thus, while it may lack generality concerning the types of TTV signals it can find, it compensates somewhat for this lack with its power to confirm the cohabitation of candidates in a system when one TTV signal is very weak---e.g., if a high-mass planet is being perturbed by a low-mass one.

Each of these methods was applied to all of the systems being considered.  The transit times for each of the analyzed candidates are produced in the manner described in \citep{Ford:2011a}.  For each system, outlier data are removed by eliminating all transit times that are four times the Median Absolute Deviation from the best fitting linear ephemeris and eliminating all transit times where the timing uncertainty is more than twice the median timing uncertainty (as was done in \citet{Ford:2012b} and \citet{Steffen:2012c}).  (The timing residuals for the sytems are shown graphically in the Appendix.
)  The results of the Monte Carlo studies, specifically the expected False Alarm Probabilities (FAP) for relevant cases with each method, are given in Table \ref{tabMC}.  In all cases at least one of the three methods generates a sufficiently low FAP to claim that the observed anticorrelation is not due to timing noise, but more likely two planet candidates in the same system.  In many cases all methods can make this claim.

% ******* Begin text from Eric

In a few cases (planets in KOIs 886, 1336, 1529, and 2086) the signal to noise ratio (SNR) of each individual transit event is relatively small.  Low SNR transits occationally result in poorly determined transit times and often these cases are ignored in general TTV studies of the \kepler\ systems (e.g., \citet{Ford:2012b,Steffen:2012c}).  In order to increase our confidence that the observed TTV signal is real, we modelled the lightcurve using a straightforward linear-plus-sinusoid ephemeris in which we allowed the times of transit to have a sinusoidal modulation.  The transit duration, impact parameter, and depth for each transiting planet were fixed (since variations in these quantities have yet to be detected for any small exoplanet).  In all cases we fit the transits of both planets simultaneously.  The results of this expanded ephemeris yielded a significant reduction in the $\chi^2$ over a simple linear ephemeris model (the reduction in $\chi^2$ being 830 for \koiees , 47 for \koiotts , 130 for \koioftn , 48 for \koitzes\ .01 and .02, and 166 for \koitzes\ .02 and .03).  These results demonstrate that the modulation in transit times, with its anticorrelation and timescale, is not due to poorly determined transit times and we proceed with our analysis in the same fashion as the other systems.

% ******** End text from Eric

\subsection{Discussion of plausible false alarms}

The choice of an FAP $<10^{-3}$ is the same as was done in \citet{Fabrycky:2012a}, \citet{Ford:2012a}, and \citet{Steffen:2012a}.  We have not published an exhaustive search for TTV systems that can be confirmed with this method as some \kepler\ systems have warranted individual scrutiny.  In addition, the vetting of all KOIs is still ongoing---even as the list of KOIs continues to grow---and therefore remains nonuniform.  Nevertheless, we do want to be confident that we are not misinterpreting these systems.  Anticorrelated TTVs with a well-defined period and phase provide a distinct functional form that yields high confidence in the pair being true interacting planets.  However, anticorrelations between parabolic TTV signatures imply a gradual change in the orbital periods, but are less powerful at eliminating alternative explanations.  We look at two possible explanations for the observed anticorrelated TTVs and show that they are significantly less likely than the single system interpretation---justifying our claim that the observed TTVs are due to planetary interactions in a single system.  First, we consider the possibility of a background star that hosts one of the planets.  Second, we consider the case of a bound stellar binary with a planet orbiting each star where the light travel time across the system produces the TTV signal.

Consider the case where both the target star and a background star host a transiting planet; and that these planets are both being perturbed by an unseen planetary companion.  Occasionally a parabolic TTV signature from one star may be anticorrelated with a parabolic TTV signal from the other.  To estimate a limit on the probability of this scenario (we do not attempt to give an exact estimate, only to constrain its value) we multiply: 1) the fraction of KOIs that show only a quadratic TTV signature, 2) the fraction of KOIs as near to first-order MMRs as those in our sample, 3) the probability of a given target having a planet candidate (a proxy for the fraction of background stars that would show a candidate), and 4) the number of candidate pairs that we investigated---then we divide by two since we require an ``anti''correlation in the TTV signals.  Recall that the resulting false alarm probability is (a limit on) the probability of misinterpreting two distinct planetary systems as a single planetary system.

Using data from \citet{Ford:2012b} we estimate the fraction of KOIs that show evidence for quadratic TTVs by selecting KOIs from \citep{Ford:2012b} where a fit using a quadratic ephemeris has a $\chi^2$ $p$-value less than 0.05.  We also require that a cubic or quartic ephemeris fail the $F$-ratio test (with a $p$-value greater than 0.1) when compared to a quadratic ephemeris---we want only systems with quadratic trends as the others are not subject to this particular misinterpretation of the TTV signal.  We find that $\sim 20\%$ of \kepler\ planet candidates show some evidence for quadratic-only trends.  Next, we note that the planet pairs we search for are within $\sim 10\%$ of a first-order MMR where we expect TTV signatures to be manifest.  The largest absolute deviation from zero, using the $\zeta_1$ statistic from \citep{Lissauer:2011b} and \citep{Fabrycky:2012b}, is 0.37.  We estimate the probability of finding two randomly chosen planet candidates having $\zeta_1<0.4$ using the distribution of KOI periods from \citep{Batalha:2012}.  We select all planet candidates with period ratios within a factor of three of the median period of all KOIs (essentially the peak of the distribution which yields a conservative estimate), calculate $\zeta_1$ relative to the median period, and divide by the total number of KOIs.  The result is that no more than 15\% of randomly chosen KOI pairs will be as close to a first order MMR as the candidates we study here.   Finally, the number of KOIs seen by \kepler\ suggests that $\sim 2\%$ of \kepler\ target stars will have a planet candidate.  Thus, no more than $(1/2) \times (0.2)^2 \times 0.15 \times 0.02 \simeq 1 \times 10^{-4}$) of randomly-selected pairs of candidates would show parabolic anticorrelations.  When identifying systems to study in this paper, we search $\sim 1000$ pairs which yields a naive expectation of 0.1 candidate pairs that show parabolic, anticorrelated TTVs.  Comparing this number to the $\sim 10$ previously confirmed systems that have have anticorrelated TTVs yields a less than 1\% chance of misinterpreting planets orbiting two different stars as a system of planets orbiting a single star.

Still, much more can be brought to bear to reduce this false alarm probability still further (not including constraints on the population of suitable background stars in the galaxy).  We look at two tests specifically.  First, and most importantly, we examined the displacement of the centroid of the target during transit.  If a pair of planets were orbiting different stars, then the centroid of the target would displace in different directions during the transits of each planet.  In all cases the displacement of the centroid is consistent with the planet pairs orbiting the same star and that the star is indeed the Kepler target.

For the second test, we calculated the quantity $\xi  \equiv (D_1/D_2)(P_2/P_1)^{1/3}$ for each planet pair where $D_1$ and $D_2$ are the transit durations and $P_1$ and $P_2$ are the orbital periods for the inner and outer planets respectively. The value of $\xi$ for each pair should be of order unity if the planets are in the same system \citep{Steffen:2010}. Indeed, the bulk of cases presented here are within the 98\% confidence interval of the distribution for observed $\xi$ values for all of the \kepler\ multiple systems. The only exception is anomalous values in the \kepstz\ (\koistz ) system (2.57 and 0.38 for the 01 -- .03 and .03 -- .02 pairs respectively). In fact, the $\xi$ value for the .03 -- .02 pair is the lowest observed to date. In this system one planet, \kepstz c (\koistz .03), appears to be on a nearly grazing orbit (see \citet{Batalha:2012})---which naturally produces such extreme values for $\xi$, however this may indicate the possibly of a false positive. As a check, we generated a synthetic population of planet canidates using a Monte Carlo code similar to \cite{Lissauer:2011b} and \cite{Fang:2012}, drawing 2 to 10 planets per system uniformly over the \kepler\ candidate period distribution and assuming circular orbits. Radii were also drawn from the empirical \kepler\ distribution.  The mutual inclinations of the planets with respect to an arbitrary reference plane were Rayleigh distributed with a mean of 1.52 degrees, equivalent to the best fit regions of \cite{Fabrycky:2012b}.  After $10^5$ realizations, $\xi$ was computed for the synthetic candidates with multiple transiting planets having total SNR $> 7.1$.  A two sample Kolmogorov–Smirnov test gives a P-value of 0.14 indicating the null-hypothesis that the synthetic and observed distributions come from the same parent distribution can not be rejected at high significance. From the synthetic population we calculate the probability of $\xi < 0.38$ or $\xi > 2.57$ to be 0.016. Given that there are 365 systems with multiple transiting planets and 529 corresponding $\xi$ pairs, a true planet system showing such extreme values is not unexpected. The $\xi$ values for the different planet pairs are also given in \ref{tabPlanets}.

The second possible misinterpretation of an anticorrelated TTV signal that we consider is planets orbiting two gravitationally-bound stars.  Here, the light travel time across the system would produce the observed period variations.  In many cases this stellar binary interpretation is not consistent with the amplitude of the TTV signal---the implied orbital period is short (a few hundred days), meaning that the distance between the stars would be small, but the TTV signal is large, requiring a large separation between the stars.  The worst-case scenario is when: 1) the binary has an orbital period equal to the duration of the data, 2) the phase of the orbit being such that the TTVs actually have the form of a cosine instead of a parabola, and 3) the second star being bright enough to allow the observed transit depth (here we use a stellar twin) but essentially massless so that the light travel time is maximized---a somewhat unphysical worst-case scenario.  None of the systems that can be modeled with sinusoids can be explained with the binary star model since their TTV amplitudes are too large.  The amplitudes of all observed parabolic TTV signals are also too large by a factor of at least 7 with the exceptions of \keptst\ (which is very near the 6:5 MMR) and \keptzes\ (which has three planets instead of just two).  Still, even these systems have amplitudes that are a factor of 2 and 1.5 too large respectively for this worst-case scenario.

In these remaining two systems (\keptst\ and \keptzes ), all of the candidate pairs are very near first order MMRs---all are comfortably within $10^{-3}$ of the resonance and have $\zeta_1 < 0.035$.  Planets this close to MMR would naturally have TTV signals that develop over very long timescales (and would be seen only as parabolas over the course of our data).  Using the same method outlined above we find that fewer than 2\% of planet pairs are this close to resonance (having $\zeta_1 < 0.04$)---and for \keptzes\ one would need two such pairs.  The probability of having a binary star system with an appropriate period ratio---between 500 and 1500 days---is approximately $0.05 / 2 = 0.025$ (using the analytic distribution given in \citet{Duquennoy:1991} and dividing by two in order to have a binary star instead of an isolated star).  Also needed is that all planets must transit their respective hosts (with a probability perhaps as large as $\sim 0.1$ given that the binary star system which would likely induce correlations in the orbital planes of the planets),  Finally, the orbital phase of the binary must be such that it produces a parabola-resembling cosine signal ($\sim 2 \times 0.2$).  Thus, the expected number of planet pairs in an plausible, stellar binary system is no larger than $0.02 \times 0.025 \times 0.1 \times 2 \times 0.2 \sim 10^{-5}$---a result slightly smaller than the scenario using a background star.  Again, this possibility is much less likely than that of a single system (nearly 1000 times less likely for these two systems where this model is potentially viable).  Nevertheless, the twin-star scenario given here has always been a known caveat for multi-transiting systems (especially for very widely separated binaries) and should such a system ever be seen, it would be a very interesting one for study.

\begin{table*}
\centering
\begin{minipage}{180mm}
\caption{Results of the Monte Carlo studies of the TTV residuals.  Only pairs with an FAP $< 0.1$ for at least one of the methods are shown.  If no realizations of the Monte Carlo were more significant than the observed signal out of $10^4$ realizations, we note the results as FAP $<10^{-3}$.  Here FAP$_1$ and FAP$_2$ come from the dynamical modelling where the inner planet is being pertubed by the outer and vice versa.  FAP$_{GP}$ comes from the Gaussian Process Monte Carlo simularion.  And, $\Xi$ and FAP$_{\Xi}$ are for the Fourier analysis Monte Carlo.  Confirmed planet pairs are noted.}
\begin{tabular}{rrrrrrrrc}
\hline
Kepler	&	KOI$_1$	&	 KOI$_2$ 	&	 FAP$_1$ 	&	 FAP$_2$ 	&	FAP$_{GP}$	&	$\Xi_{\text{max}}$	&	FAP$_\Xi$	&	Confirmed	\\	\hline \hline
48	&	148.01	&	148.02	&	0.0022	&	$<10^{-3}$	&		&	19.6	&	$<10^{-3}$	&	y	\\	
	&	148.01	&	148.03	&		&		&		&		&		&		\\	
	&	148.02	&	148.03	&		&		&		&		&		&		\\	\hline
49	&	248.03	&	248.01	&		&		&		&		&		&		\\	
	&	248.03	&	248.02	&		&		&		&		&		&		\\	
	&	248.03	&	248.04	&		&	0.0816	&		&		&		&		\\	
	&	248.01	&	248.02	&	$<10^{-3}$	&	$<10^{-3}$	&		&	69.8	&	$<10^{-3}$	&	y	\\	
	&	248.01	&	248.04	&		&		&		&	7.1	&		&		\\	
	&	248.02	&	248.04	&		&		&		&		&		&		\\	\hline
50	&	262.01	&	262.02	&	$<10^{-3}$	&	0.0120	&	$<10^{-3}$	&	34.8	&	$<10^{-3}$	&	y	\\	\hline
51	&	620.01	&	620.03	&	0.0001	&	0.0166	&	0.0006	&	79.6	&	0.0072	&	y	\\	
	&	620.01	&	620.02	&		&		&		&		&		&		\\	
	&	620.03	&	620.02	&	0.0114	&		&		&		&		&		\\	\hline
52	&	775.02	&	775.01	&	0.0003	&	0.0476	&		&	19.1	&	0.0001	&	y	\\	
	&	775.02	&	775.03	&		&		&		&		&		&		\\	
	&	775.01	&	775.03	&	0.0439	&		&		&		&		&		\\	\hline
53	&	829.02	&	829.01	&		&		&		&	5.9	&		&		\\	
	&	829.02	&	829.03	&		&	0.0019	&	0.0004	&		&		&		\\	
	&	829.01	&	829.03	&	0.0009	&	0.0007	&		&	10.9	&	0.0072	&	y	\\	\hline
54	&	886.01	&	886.02	&	$<10^{-3}$	&	$<10^{-3}$	&	$<10^{-3}$	&	104.3	&	$<10^{-3}$	&	y	\\	
	&	886.01	&	886.03	&		&		&		&		&		&		\\	
	&	886.02	&	886.03	&		&		&		&	6.0	&		&		\\	\hline
55	&	904.01	&	904.04	&		&		&		&	7.7	&		&		\\	
	&	904.01	&	904.05	&		&		&		&		&		&		\\	
	&	904.01	&	904.02	&		&		&		&		&		&		\\	
	&	904.01	&	904.03	&		&	$<10^{-3}$	&		&		&		&		\\	
	&	904.04	&	904.05	&		&		&		&		&		&		\\	
	&	904.04	&	904.02	&		&	0.0001	&		&		&		&		\\	
	&	904.04	&	904.03	&		&		&		&	6.9	&		&		\\	
	&	904.05	&	904.02	&		&		&		&	6.2	&		&		\\	
	&	904.05	&	904.03	&		&		&		&		&		&		\\	
	&	904.02	&	904.03	&	$<10^{-3}$	&	$<10^{-3}$	&		&	93.2	&	$<10^{-3}$	&	y	\\	\hline
56	&	1241.02	&	1241.01	&	$<10^{-3}$	&	0.0210	&		&	11.5	&	0.0016	&	y	\\	\hline
57	&	1270.01	&	1270.02	&		&	$<10^{-3}$	&		&	18.1	&	$<10^{-3}$	&	y	\\	\hline
58	&	1336.01	&	1336.02	&	0.0001	&	0.0001	&		&	13.6	&	0.0006	&	y	\\	\hline
59	&	1529.02	&	1529.01	&		&	$<10^{-3}$	&		&		&	0.28	&	y	\\	\hline
60	&	2086.01	&	2086.02	&	0.0002	&	0.0092	&		&	10.0	&	0.0049	&	y	\\	
	&	2086.01	&	2086.03	&		&		&	0.0001	&	7.8	&		&	y	\\	
	&	2086.02	&	2086.03	&	0.0194	&	0.0001	&		&		&		&	y	\\	\hline \hline
\end{tabular}
\label{tabMC}
\end{minipage}
\end{table*}

\subsection{Mass limits from orbital stability}

Once the fact that the different planet candidates orbit the same star has been established, we use dynamical stability to show that their masses must be planetary.  We followed the method used and described in \cite{Fabrycky:2012a}.  Briefly, we integrate a series of planetary systems with different masses, starting from the value at which stability is analytically assured \citep{Gladman:1993}.  We integrate the system until it becomes unstable, produces close encounters, or survives without close encounters for $10^9$ inner-planet orbits.  We set the upper limit on the planet mass as the smallest mass that yields an unstable system.  For each planet candidate examined with the stability test, this value is $\lesssim 25 M_\text{Jupiter}$, confirming them as planets (using the planet criterion suggested by \citep{Schneider:2011}).  Table \ref{tabMass} shows the overall results of the stability study.

\citet{Ford:2012a} discusses in detail the possible, though unlikely, scenario where the planets in those systems are both orbiting a background star instead of the target star.  We mention here the implications for the planet nature of the orbiting objects.  Dynamical stability yields constraints on the masses of the planets compared to the total mass of the system.  So, if a background star is more massive than the target, then the corresponding mass limits would grow in proportion to the difference.  It is possible that such a scenario would push the mass limits beyond the planetary regime.  For most systems the mass of the true host star would need to be larger by a factor of 10 or more (for most of the systems here that would imply $>8M_\odot$ host).  Such massive stars are quite rare and don't survive on the main sequence for very long.  For the less massive targets with large mass limits on the orbiting objects---\kepssf\ (\koissf) in particular---a more massive background star, with a factor of two or more greater mass, could be an issue (though the orbits would all need to be oriented in a physically unfavorable manner to yield consistent transit durations).  Nevertheless, refined mass limits from the next section effectively eliminate the possibility that the objects orbiting \kepssf\ are not planetary.

\begin{table}
\centering
\caption{Mass limits from the stability analysis.}
\begin{tabular}{lrr}
\hline
Planet	&	KOI				&	Max Mass ($M_{\text{J}}$)	\\	\hline	\hline
48b	&	148.	01	&	5.94	\\		
48c	&	148.	02	&	11.61	\\		
49b	&	248.	01	&	0.98	\\		
49c	&	248.	02	&	0.72	\\		
50b	&	262.	01	&	0.10	\\		
50c	&	262.	02	&	0.11	\\		
51b	&	620.	01	&	3.23	\\		
51c	&	620.	03	&	2.60	\\		
52b	&	775.	02	&	8.70	\\		
52c	&	775.	01	&	10.41	\\		
53b	&	829.	01	&	18.41	\\		
53c	&	829.	03	&	15.74	\\		
54b	&	886.	01	&	0.92	\\		
54c	&	886.	02	&	0.37	\\		
55b	&	904.	02	&	1.49	\\		
55c	&	904.	03	&	1.11	\\		
56b	&	1241.02	&	5.12	\\		
56c	&	1241.01	&	12.18	\\		
57b	&	1270.01	&	18.86	\\		
57c	&	1270.02	&	6.95	\\		
58b	&	1336.01	&	1.39	\\		
58c	&	1336.02	&	2.19	\\		
59b	&	1529.02	&	2.05	\\		
59c	&	1529.01	&	1.37	\\		
60b	&	2086.01	&	0.25	\\		
60c	&	2086.02	&	0.56	\\		
60d	&	2086.03	&	0.68	\\ \hline
\end{tabular}
\label{tabMass}
\end{table}

\subsection{Analytic mass limits}

A recent paper by \citet{Lithwick:2012b} gives analytic formulae that can be used to derive mass estimates or upper limits using the relative amplitudes and phases of the TTV signal.  Those formulae apply once a reliable amplitude for a sinusoid-shaped TTV signal can be extracted.  Here we assume that the free eccentricity is zero and apply the corresponding formula from \citep{Lithwick:2012b} to the systems where this analysis is appropriate (i.e., where the amplitude of the TTV signal can be extracted).

The free eccentricity of a planet---the eccentricity it would have in isolation (as opposed to the forced eccentricity, which results from the presence of the perturbing planet)---will tend to boost the TTV signal on a nearby planet.  Thus, if there is some free eccentricity, then the mass estimates that come from the above analysis should be interpreted as approximate mass upper limits.  A constraint on the free eccentricity can be made from the phase of the TTV signal relative to the line of sight of the observer.  In our sample of systems the uncertainty in the TTV phase is sufficiently large that we do not claim absence of free eccentricity in any system.  Thus, we consider all estimates to be upper limits.  Nevertheless, these upper limits are roughly one order of magnitude more stringent than what one obtains from dynamical stability.

The results of this analysis are presented in Table \ref{tabAnalyticmass} where we include the mass limits from orbital stability (copied from table \ref{tabMass} but now in Earth masses), the analytic mass estimate assuming zero free eccentricity, the formal uncertainty in this estimate, and the measured TTV phase and its uncertainty (note that we quote the best fitting value for the masses, so the actual mass limit at a specific confidence level must be calculated from this quantity and its given uncertainty).  The systems analyzed in this manner include \kepofe , \keptfe , \kepssf , \kepetn , \kepotsz , and \kepotts .  For these systems, the location in time where the line of conjunction is along the line of sight is shown in the corresponding figures in the Appendix.  Zero TTV phase (measured for the inner planet) occurs when the TTV signal crosses zero from above.    A more detailed study of these systems and those discussed in \citep{Lithwick:2012b} will yield an interesting estimate for the number of near-resonant systems with free eccentricity.  At first glance, it appears that rougly half of the systems studied using this approach retain some free eccentricity ($\sim$ 6 out of 12, though more data will inevitably yield more a more reliable count).  Such a result will have implications for how the excess of planetary systems near, but perhaps not in, low-order MMRs is generated (\citep{Lithwick:2012a,Terquem:2007,Papaloizou:2011,Podlewska:2012,Sandor:2006,Batygin:2012}).

\begin{table*}
\centering
\caption{Mass limits from orbital stability (now given in Earth masses) and from analytic formulae with TTV phase $\phi$.  The quoted numbers from the analytic results are the best fitting values, so the actual mass limits will be a combination of these numbers and their uncertainties.}
\begin{tabular}{lrccccc}
\hline
Kepler 		& KOI 	& Stability Mass Limit  & TTV Mass Limit & $\sigma_M$ & $\phi$ (deg) & $\sigma_\phi$ \\ 
& & ($M_\oplus$) & ($M_\oplus$) & ($M_\oplus$) & (degrees) & (degrees) \\ \hline \hline
48b &   148.01 	& 1960	& $17.2$ 	& 3.9 	& -7.9 	& 14 	\\
48c &   148.02 	& 3830	& $10.1$ 	& 3.5 	& -156 	& 10 	\\
50b &   248.01 	& 323	& $7.6$	& 1.3 	& 47 	& 9.0 	\\
50c &   248.02 	& 238	& $7.1$ 	& 1.1 	& -138 	& 9.3 	\\
52b &   775.02 	& 2870	& $89$ 	& 46 	& -153 	& 9.6 	\\
52c &   775.01 	& 3440	& $37.4$ 	& 8.0 	& 39 	& 27 	\\
53b &   829.01 	& 6080	& $84$ 	& 27 	& -42 	& 26 	\\
53c &   829.03 	& 5190	& $24$ 	& 12 	& 111 	& 40 	\\
57b & 1270.01 	& 6220	& $100$ 	& 15 	& 98 	& 190 	\\
57c & 1270.02 	& 2290	& $5.4$ 	& 3.7 	& -84 	& 60		\\
58b & 1336.01 	& 677	& $27.4$ 	& 8.1 	& -85 	& 122 	\\
58c & 1336.02 	& 452	& $41$ 	& 12 	& 135 	& 17 	\\ \hline
\end{tabular}
\label{tabAnalyticmass}
\end{table*}

\section{Discussion and Conclusion}\label{secConclusion}

All of the systems that we confirm here and in similar previous studies are near first-order MMR.  In this case there are six pairs near the 2:1 MMR (\kepofe , \kepstz , \kepssf , \kepetn , \kepotfo , and \kepotsz ), five pairs near the 3:2 MMR (\keptfe , \kepees , \kepnzf , \kepotts , and \kepoftn ), and one pair each in the 4:3 (\keptzes\ c and d ), 5:4 (\keptzes\ b and c), and 6:5 (\keptst ).  Several systems show additional planet candidates that, due to their presence in known multi-transiting systems, are very likely to be true planets as well \citep{Lissauer:2012}.  If we consider all of the KOIs to be planets, then a few systems will have multiple pairs of planets near first order MMRs with some of these pairs being adjacent links in a near resonant chain including: \kepstz\ near a 3:2 -- 2:1 chain (3:2:1), \kepetn\ near a 2:1 -- 2:1 chain (4:2:1), \kepnzf\ with one pair near the 2:1 and a separate pair near the 3:2, and \keptzes\ near a 5:4 -- 4:3 chain (20:15:12).  The balance of the planet pairs in these systems, if near a resonance at all, must be near a resonance of higher order.

These planetary systems, like the other systems that have been confirmed via TTV analyses, are of particular interest for long-term scrutiny with follow-up observations and dynamical studies.  The detailed dynamics of planetary systems yield meaningful information about the evolutionary histories of the orbital architectures of the system.  For example, the differences in the fraction of systems that show multiple candidates and detectabe TTVs among smaller planet candidates with few-day orbital periods compared with the larger, Jupiter-sized candidates in the same period range---these hot Jupiters have no compelling signs of current planet-planet dynamics---gives strong clues as to their likely distinct dynamical history compared to the bulk of the exoplanet population \citep{Steffen:2012b}.

These newly confirmed planetary systems continue to show the value that transit timing variations have and will have in transiting planet endeavors.  As new planets are found, with increasingly smaller sizes and longer orbital periods, TTVs provide the only viable means of determining planetary masses.  TTVs, along with statistical validation via the techniques employed in \citep{Torres:2011} and \citep{Lissauer:2012}, will likely be the primary means to demonstrate the planetary nature of the planet candidates that the \kepler\ spacecraft identifies (in part because the stars in the \kepler\ field are often too dim to make good Radial Velocity targets).

Two of these systems show a modest gap in the orbital period ratios of adjacent pairs of KOIs.  If planetary systems tend to be dynamically packed (e.g. \citet{Barnes:2004,Lissauer:1995,Laskar:2000}) then there may be additional undetected planets in these gaps.  In this set of systems, the most promising examples for future investigations include \kepofe\ which has a gap that is over a factor of 4 in period ratio and \koitfe\ which has a gap that is nearly a factor of three in orbital period.  A detailed TTV analysis of these two systems would provide useful constraints on or support for the hypothesis of dynamically packed systems.

The planetary system orbiting the giant star \kepotfo\ is another interesting individual system---being an early discovery of a transiting multiplanet system hosted by a giant.  Chromospheric emission from giant stars causes the star to be limb brightened in certain pass bands (e.g. the Calcium H and K lines).  Comparison of the transit duration in these bands with the duration from the broad band Kepler data may yield a direct measurement of the size of the chromosphere \citep{Assef:2009}.

Continued monitoring by the \kepler\ spacecraft and future study will likely yield more precise mass measurements of the planets in these systems and consequently provide constraints on their bulk composition.  Such mass measurements will in turn improve our understanding of planet formation processes outside of our own solar system and are useful to place or solar system in the context of the general planet population.

\section*{Acknowledgements}
We thank Yoram Lithwick, Roberto Sanchis-Ojeda, Bill Chaplin, and Daniel Huber for their useful input and discussions.  Funding for the \kepler\ mission is provided by NASA's Science Mission Directorate.  We thank the entire Kepler team for the many years of work that is proving so successful.  J.H.S acknowledges support by NASA under grant NNX08AR04G issued through the Kepler Participating Scientist Program.  D. C. F. and J. A. C. acknowledge support for this work was provided by NASA through Hubble Fellowship grants \#HF-51272.01-A and \#HF-51267.01-A awarded by the Space Telescope Science Institute, which is operated by the Association of Universities for Research in Astronomy, Inc., for NASA, under contract NAS 5-26555.  R.C.M. is supported by the National Science Foundation Graduate Research Fellowship under Grant No. DGE-0802270.

\bibliographystyle{plainnat}
%\bibliography{multis}

\begin{thebibliography}{36}
\providecommand{\natexlab}[1]{#1}
\providecommand{\url}[1]{\texttt{#1}}
\expandafter\ifx\csname urlstyle\endcsname\relax
  \providecommand{\doi}[1]{doi: #1}\else
  \providecommand{\doi}{doi: \begingroup \urlstyle{rm}\Url}\fi

\bibitem[{Agol} et~al.(2005){Agol}, {Steffen}, {Sari}, and
  {Clarkson}]{Agol:2005}
E.~{Agol}, J.~{Steffen}, R.~{Sari}, and W.~{Clarkson}.
\newblock \emph{\mnras}, 359:\penalty0 567--579, May 2005.

\bibitem[{Assef} et~al.(2009){Assef}, {Gaudi}, and {Stanek}]{Assef:2009}
R.~J. {Assef}, B.~S. {Gaudi}, and K.~Z. {Stanek}.
\newblock \emph{\apj}, 701:\penalty0 1616--1626, August 2009.

\bibitem[{Barnes} and {Raymond}(2004)]{Barnes:2004}
R.~{Barnes} and S.~N. {Raymond}.
\newblock \emph{\apj}, 617:\penalty0 569--574, December 2004.

\bibitem[{Batalha} et~al.(2012)]{Batalha:2012}
N.~M. {Batalha} et~al.
\newblock \emph{ArXiv e-prints}, February 2012.

\bibitem[{Batygin} and {Morbidelli}(2012)]{Batygin:2012}
K.~{Batygin} and A.~{Morbidelli}.
\newblock \emph{ArXiv e-prints:1204.2791}, 2012.

\bibitem[{Borucki} et~al.(2011)]{Borucki:2011}
W.~J. {Borucki} et~al.
\newblock \emph{\apj}, 736:\penalty0 19--+, July 2011.

\bibitem[{Carter} et~al.(2012){Carter}, {Agol}, et~al.]{Carter:2012}
J.~A. {Carter}, E.~{Agol}, et~al.
\newblock \emph{ArXiv e-prints:1206.4718}, June 2012.

\bibitem[{Cochran} et~al.(2011)]{Cochran:2011}
W.~D. {Cochran} et~al.
\newblock \emph{\apjs}, 197:\penalty0 7, November 2011.

\bibitem[{Demarque} et~al.(2004){Demarque}, {Woo}, {Kim}, and
  {Yi}]{Demarque:2004}
P.~{Demarque}, J.-H. {Woo}, Y.-C. {Kim}, and S.~K. {Yi}.
\newblock \emph{\apjs}, 155:\penalty0 667--674, December 2004.

\bibitem[{Duquennoy} and {Mayor}(1991)]{Duquennoy:1991}
A.~{Duquennoy} and M.~{Mayor}.
\newblock \emph{\aap}, 248:\penalty0 485--524, August 1991.

\bibitem[{Fabrycky} et~al.(2012{\natexlab{a}}){Fabrycky}, {Ford}, {Steffen},
  et~al.]{Fabrycky:2012a}
D.~C. {Fabrycky}, E.~B. {Ford}, J.~H. {Steffen}, et~al.
\newblock \emph{\apj}, 750:\penalty0 114, May 2012{\natexlab{a}}.

\bibitem[{Fabrycky} et~al.(2012{\natexlab{b}})]{Fabrycky:2012b}
D.~C. {Fabrycky} et~al.
\newblock \emph{ArXiv e-prints:1202.6328}, February 2012{\natexlab{b}}.

\bibitem[{Fang} and {Margot}(2012)]{Fang:2012}
J.~{Fang} and J.-L. {Margot}.
\newblock \emph{ArXiv e-prints}, July 2012.

\bibitem[{Ford} et~al.(2012{\natexlab{a}}){Ford}, {Fabrycky}, {Steffen},
  et~al.]{Ford:2012a}
E.~B. {Ford}, D.~C. {Fabrycky}, J.~H. {Steffen}, et~al.
\newblock \emph{\apj}, 750:\penalty0 113, May 2012{\natexlab{a}}.

\bibitem[{Ford} et~al.(2011)]{Ford:2011a}
E.~B. {Ford} et~al.
\newblock \emph{\apjs}, 197:\penalty0 2--+, November 2011.

\bibitem[{Ford} et~al.(2012{\natexlab{b}})]{Ford:2012b}
E.~B. {Ford} et~al.
\newblock \emph{ArXiv e-prints:1201.1892}, January 2012{\natexlab{b}}.

\bibitem[{Gladman}(1993)]{Gladman:1993}
B.~{Gladman}.
\newblock \emph{\icarus}, 106:\penalty0 247, November 1993.

\bibitem[{Holman} and {Murray}(2005)]{Holman:2005}
M.~J. {Holman} and N.~W. {Murray}.
\newblock \emph{Science}, 307:\penalty0 1288--1291, February 2005.

\bibitem[{Laskar}(2000)]{Laskar:2000}
J.~{Laskar}.
\newblock \emph{Physical Review Letters}, 84:\penalty0 3240--3243, April 2000.

\bibitem[{Lissauer}(1995)]{Lissauer:1995}
J.~J. {Lissauer}.
\newblock {Urey prize lecture: On the diversity of plausible planetary
  systems}.
\newblock \emph{\icarus}, 114:\penalty0 217--236, April 1995.

\bibitem[{Lissauer} et~al.(2011{\natexlab{a}})]{Lissauer:2011a}
J.~J. {Lissauer} et~al.
\newblock \emph{Nature}, 470:\penalty0 53, January 2011{\natexlab{a}}.

\bibitem[{Lissauer} et~al.(2011{\natexlab{b}})]{Lissauer:2011b}
J.~J. {Lissauer} et~al.
\newblock \emph{\apjs}, 197:\penalty0 8, November 2011{\natexlab{b}}.

\bibitem[{Lissauer} et~al.(2012)]{Lissauer:2012}
J.~J. {Lissauer} et~al.
\newblock \emph{\apj}, 750:\penalty0 112, May 2012.

\bibitem[{Lithwick} and {Wu}(2012)]{Lithwick:2012a}
Y.~{Lithwick} and Y.~{Wu}.
\newblock \emph{ArXiv e-prints:1204.2555}, April 2012.

\bibitem[{Lithwick} et~al.(2012){Lithwick}, {Xie}, and {Wu}]{Lithwick:2012b}
Y.~{Lithwick}, J.~{Xie}, and Y.~{Wu}.
\newblock \emph{ArXiv e-prints:1207.4192}, July 2012.

\bibitem[Lomb(1976)]{Lomb:1976}
N.R. Lomb.
\newblock \emph{Astrophys.Space Sci.}, 39:\penalty0 447--462, 1976.

\bibitem[{Papaloizou}(2011)]{Papaloizou:2011}
J.~C.~B. {Papaloizou}.
\newblock \emph{Celestial Mechanics and Dynamical Astronomy}, 111:\penalty0
  83--103, October 2011.

\bibitem[{Podlewska-Gaca} et~al.(2012){Podlewska-Gaca}, {Papaloizou}, and
  {Szuszkiewicz}]{Podlewska:2012}
E.~{Podlewska-Gaca}, J.~C.~B. {Papaloizou}, and E.~{Szuszkiewicz}.
\newblock \emph{\mnras}, 421:\penalty0 1736--1756, April 2012.

\bibitem[{S{\'a}ndor} and {Kley}(2006)]{Sandor:2006}
Z.~{S{\'a}ndor} and W.~{Kley}.
\newblock \emph{\aap}, 451:\penalty0 L31--L34, June 2006.

\bibitem[{Schneider} et~al.(2011){Schneider}, {Dedieu}, {Le Sidaner},
  {Savalle}, and {Zolotukhin}]{Schneider:2011}
J.~{Schneider}, C.~{Dedieu}, P.~{Le Sidaner}, R.~{Savalle}, and
  I.~{Zolotukhin}.
\newblock \emph{\aap}, 532:\penalty0 A79, August 2011.

\bibitem[{Steffen} et~al.(2012{\natexlab{a}}){Steffen}, {Fabrycky}, {Ford},
  et~al.]{Steffen:2012a}
J.~H. {Steffen}, D.~C. {Fabrycky}, E.~B. {Ford}, et~al.
\newblock \emph{\mnras}, 421:\penalty0 2342--2354, April 2012{\natexlab{a}}.

\bibitem[{Steffen} et~al.(2012{\natexlab{b}}){Steffen}, {Ford},
  et~al.]{Steffen:2012c}
J.~H. {Steffen}, E.~B. {Ford}, et~al.
\newblock \emph{ArXiv e-prints:1201.1873}, January 2012{\natexlab{b}}.

\bibitem[{Steffen} et~al.(2010)]{Steffen:2010}
J.~H. {Steffen} et~al.
\newblock \emph{\apj}, 725:\penalty0 1226--1241, December 2010.

\bibitem[Steffen et~al.(2012)]{Steffen:2012b}
Jason~H. Steffen et~al.
\newblock \emph{Proc.Nat.Acad.Sci.}, 109:\penalty0 7982--7987, 2012.
\newblock \doi{10.1073/pnas.1120970109}.

\bibitem[{Terquem} and {Papaloizou}(2007)]{Terquem:2007}
C.~{Terquem} and J.~C.~B. {Papaloizou}.
\newblock {Migration and the Formation of Systems of Hot Super-Earths and
  Neptunes}.
\newblock \emph{\apj}, 654:\penalty0 1110--1120, January 2007.

\bibitem[{Torres} et~al.(2011)]{Torres:2011}
G.~{Torres} et~al.
\newblock \emph{\apj}, 727:\penalty0 24, January 2011.

\end{thebibliography}

\appendix

\section{Timing residuals for each system.\label{secResiduals}}

\begin{figure}
\includegraphics[width=0.45\textwidth]{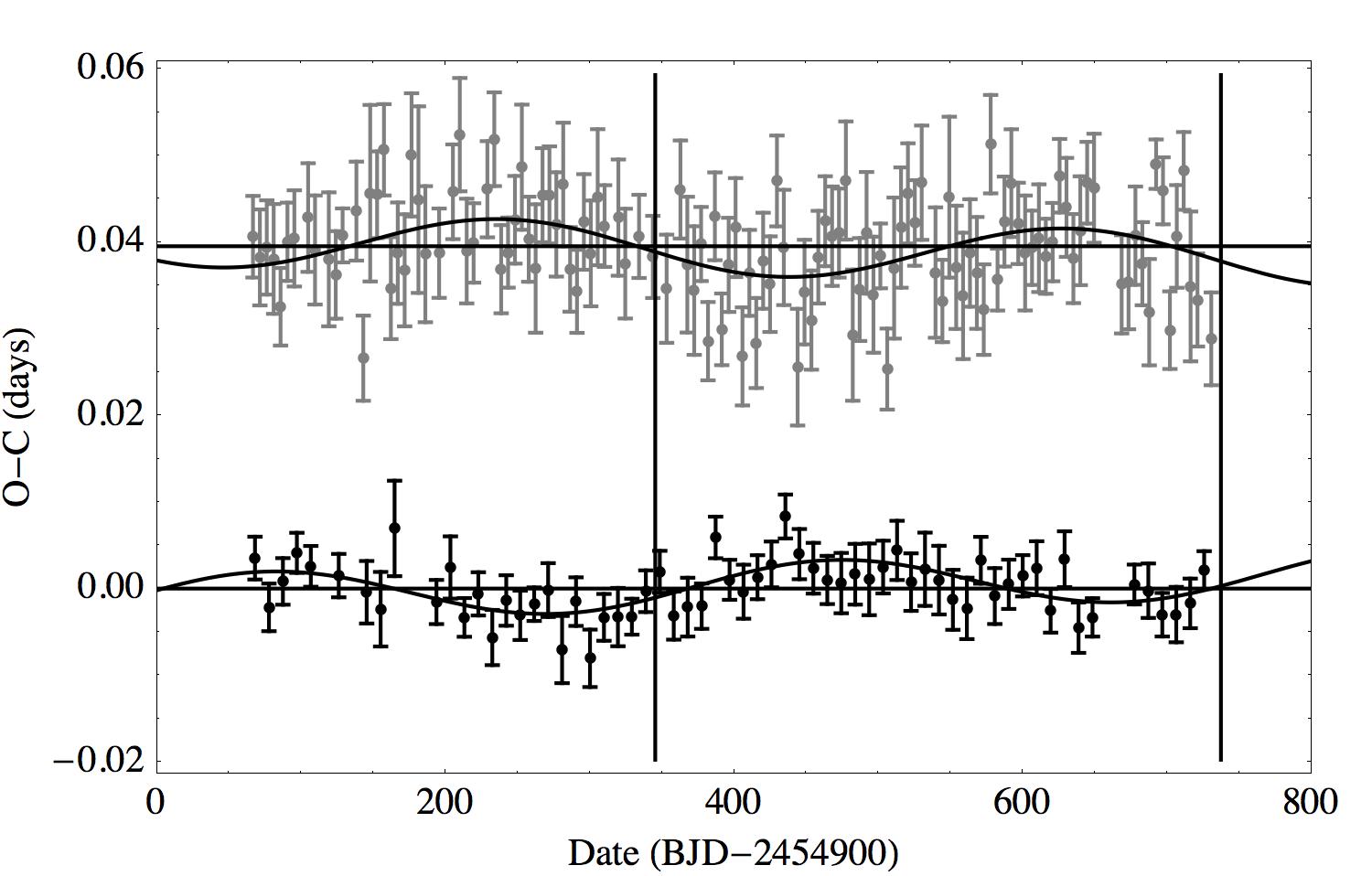}
\caption{Deviations from a constant period for \kepofe\ (\koiofe).  The residuals for the outer planet have been displaced vertically for convenience in seeing the TTV signal.  Vertical lines correspond to the times when the line of conjunctions of the two planets crosses the line of sight and are used to measure the TTV phase for the analytic mass estimates shown in Table \ref{tabAnalyticmass}.}
\label{koi0148}
\end{figure}

\begin{figure}
\includegraphics[width=0.45\textwidth]{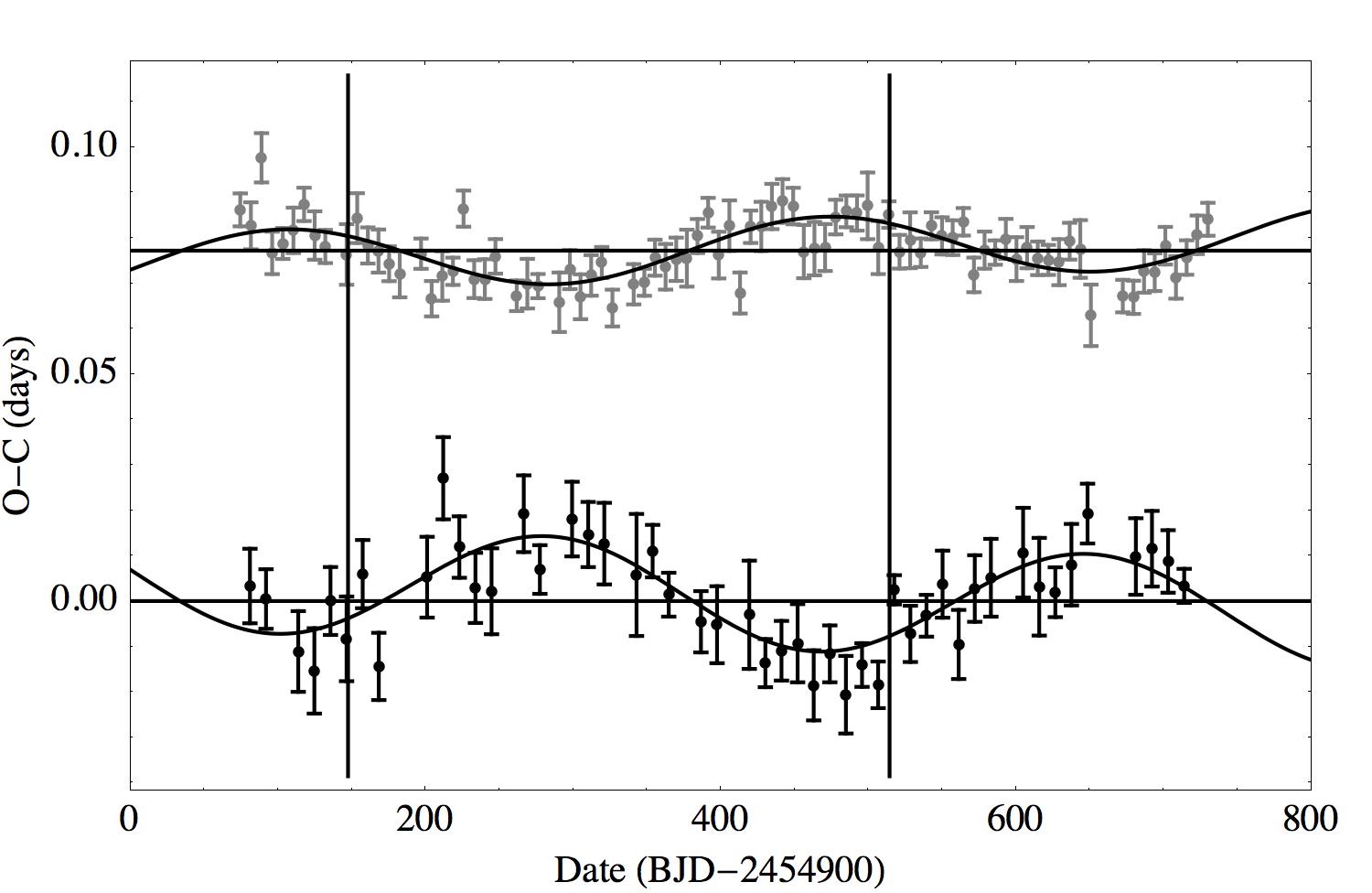}
\caption{Deviations from a constant period for \keptfe\ (\koitfe).  The residuals for the outer planet have been displaced vertically for convenience in seeing the TTV signal.  Vertical lines correspond to the times when the line of conjunctions of the two planets crosses the line of sight and are used to measure the TTV phase for the analytic mass estimates shown in Table \ref{tabAnalyticmass}.}
\label{koi0248}
\end{figure}

\begin{figure}
\includegraphics[width=0.45\textwidth]{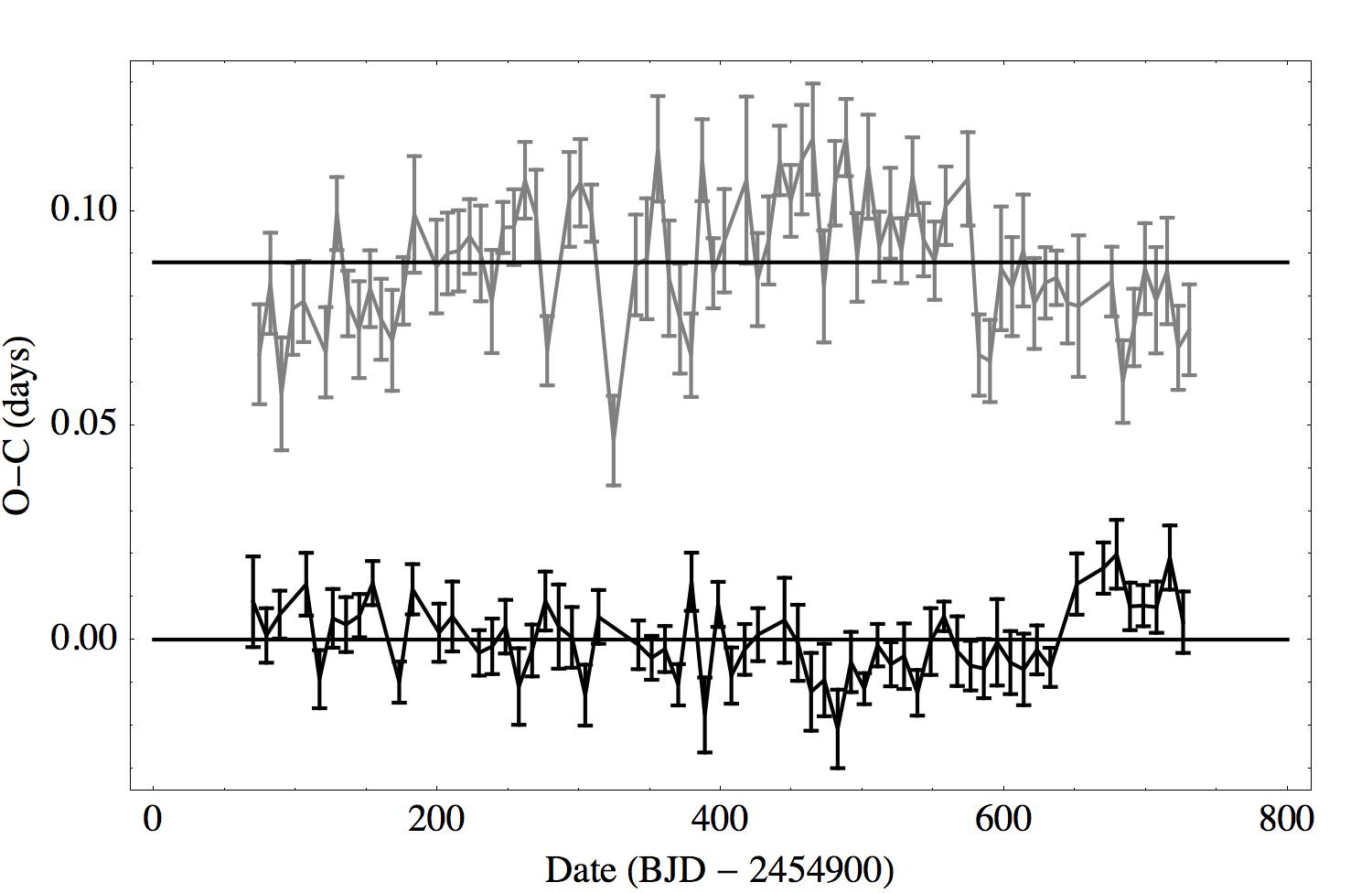}
\caption{Deviations from a constant period for \keptst\ (\koitst).  The residuals for the outer planet have been displaced vertically for convenience in seeing the TTV signal.}
\label{koi0262}
\end{figure}

\begin{figure}
\includegraphics[width=0.45\textwidth]{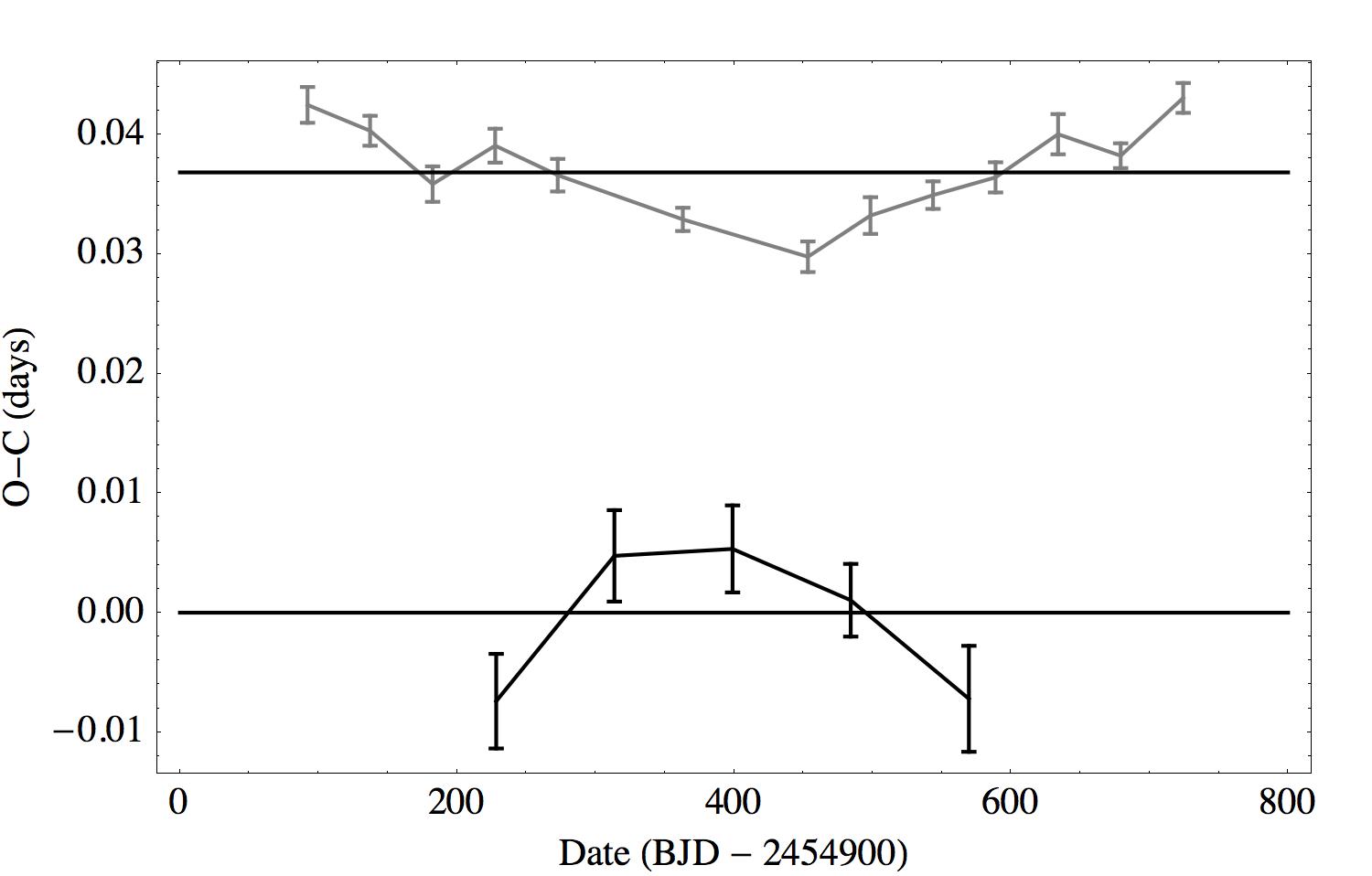}
\caption{Deviations from a constant period for \kepstz\ (\koistz).  The residuals for the outer planet have been displaced vertically for convenience in seeing the TTV signal.}
\label{koi0620}
\end{figure}

\begin{figure}
\includegraphics[width=0.45\textwidth]{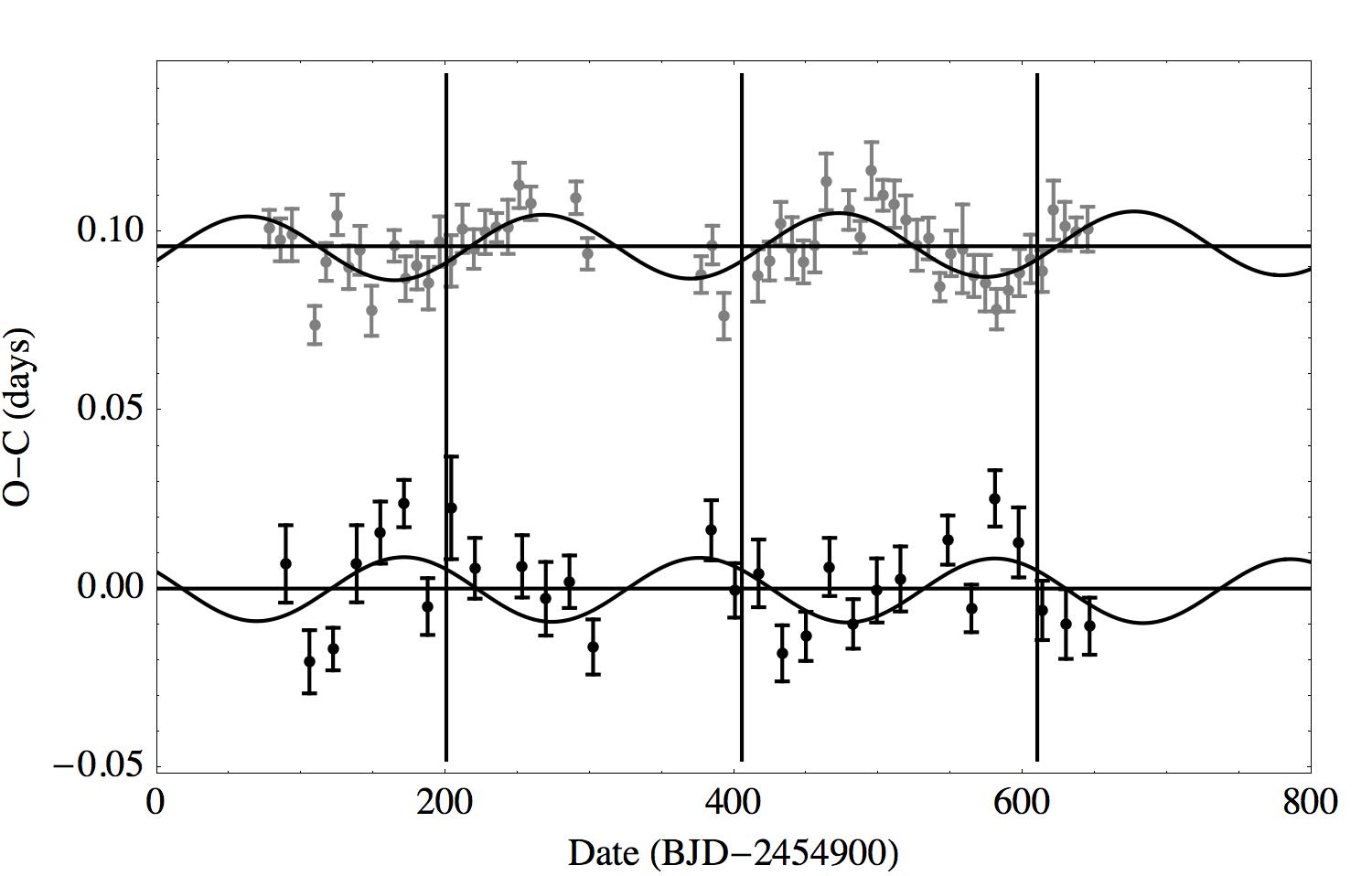}
\caption{Deviations from a constant period for \kepssf\ (\koissf).  The residuals for the outer planet have been displaced vertically for convenience in seeing the TTV signal.  Vertical lines correspond to the times when the line of conjunctions of the two planets crosses the line of sight and are used to measure the TTV phase for the analytic mass estimates shown in Table \ref{tabAnalyticmass}.}
\label{koi0775}
\end{figure}

\begin{figure}
\includegraphics[width=0.45\textwidth]{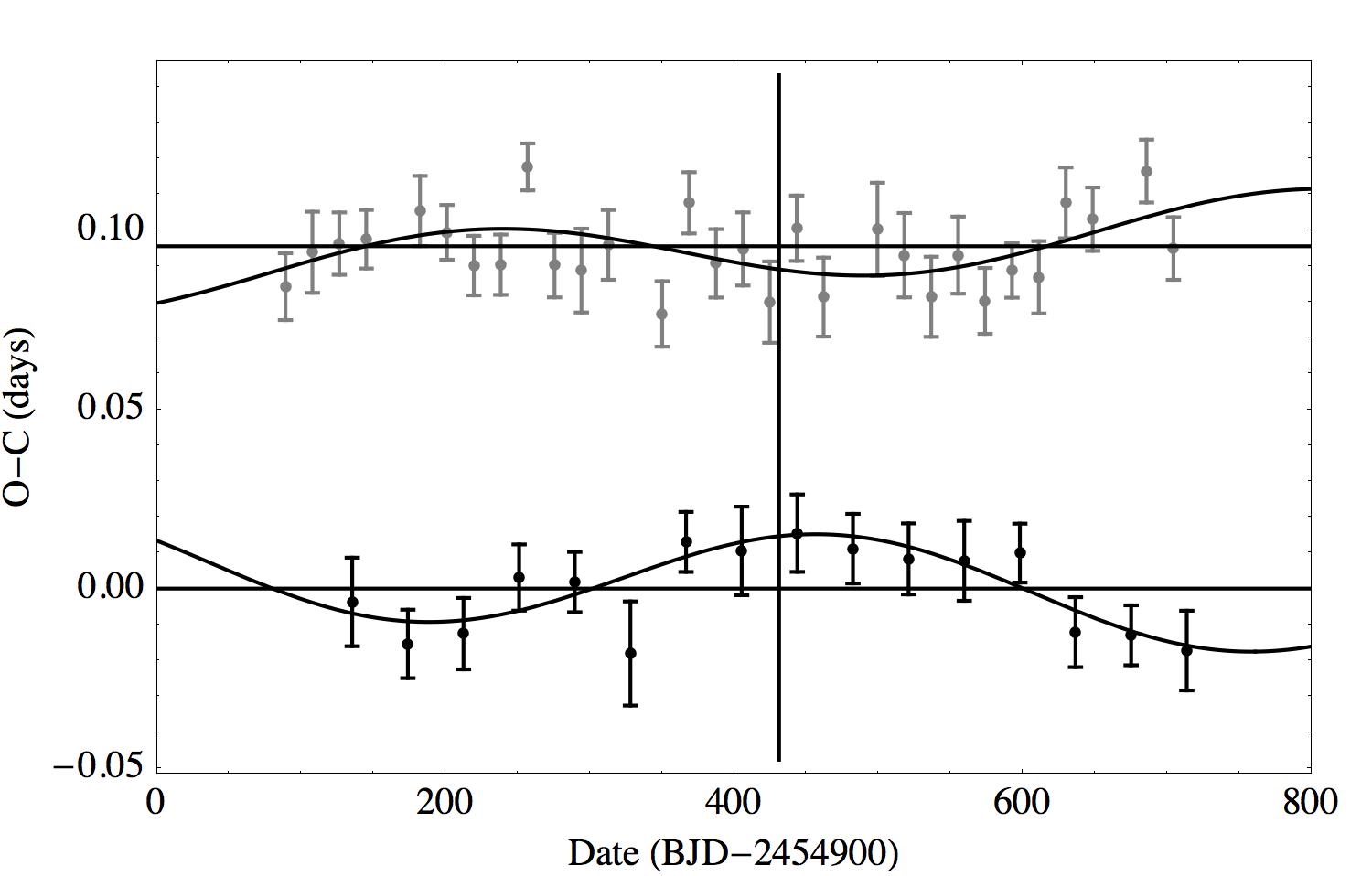}
\caption{Deviations from a constant period for \kepetn\ (\koietn).  The residuals for the outer planet have been displaced vertically for convenience in seeing the TTV signal.  Vertical lines correspond to the times when the line of conjunctions of the two planets crosses the line of sight and are used to measure the TTV phase for the analytic mass estimates shown in Table \ref{tabAnalyticmass}.}
\label{koi0829}
\end{figure}

\begin{figure}
\includegraphics[width=0.45\textwidth]{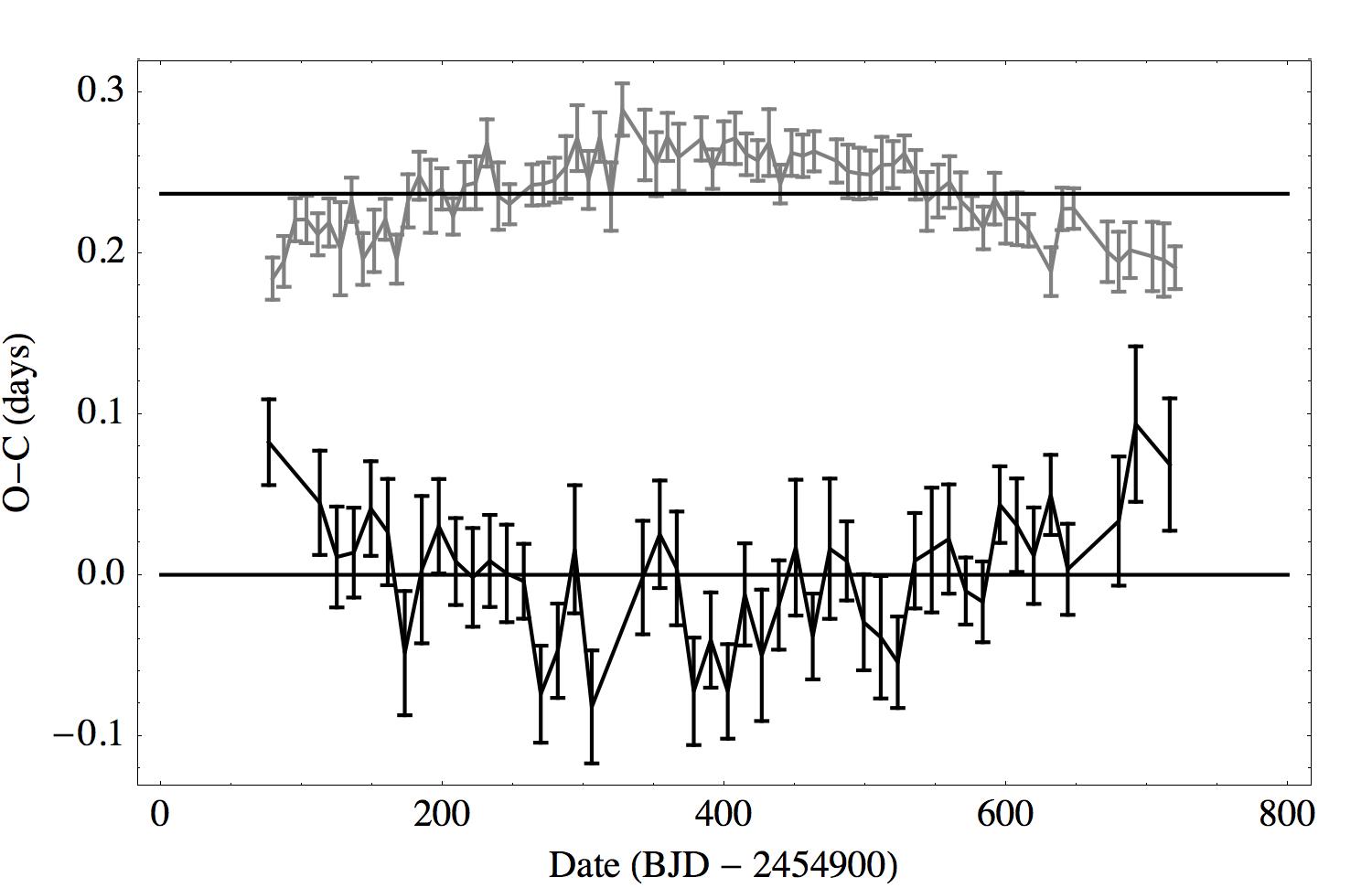}
\caption{Deviations from a constant period for \kepees\ (\koiees).  The residuals for the outer planet have been displaced vertically for convenience in seeing the TTV signal.}
\label{koi0886}
\end{figure}

\begin{figure}
\includegraphics[width=0.45\textwidth]{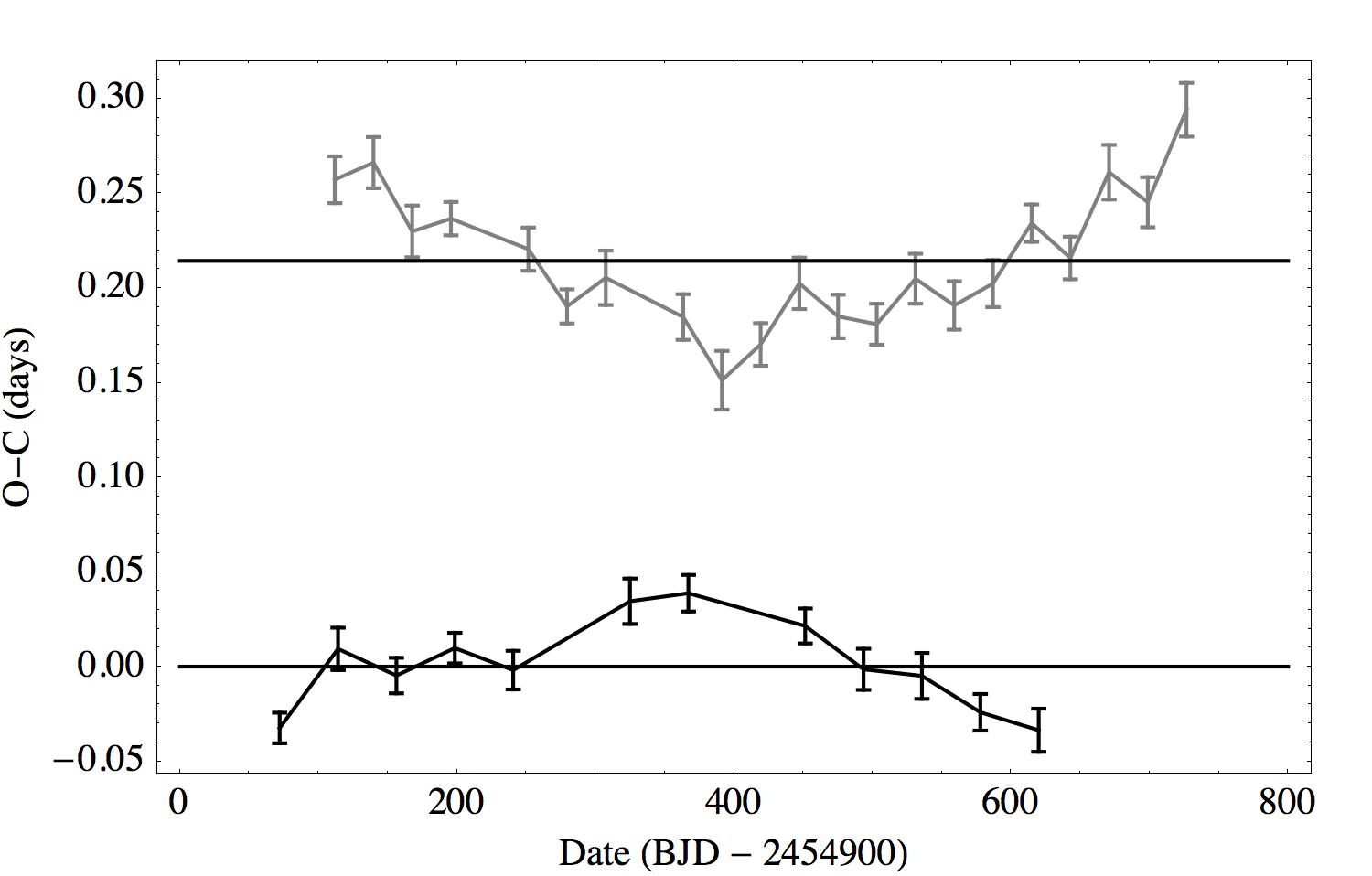}
\caption{Deviations from a constant period for \kepnzf\ (\koinzf).  The residuals for the outer planet have been displaced vertically for convenience in seeing the TTV signal.}
\label{koi0904}
\end{figure}

\begin{figure}
\includegraphics[width=0.45\textwidth]{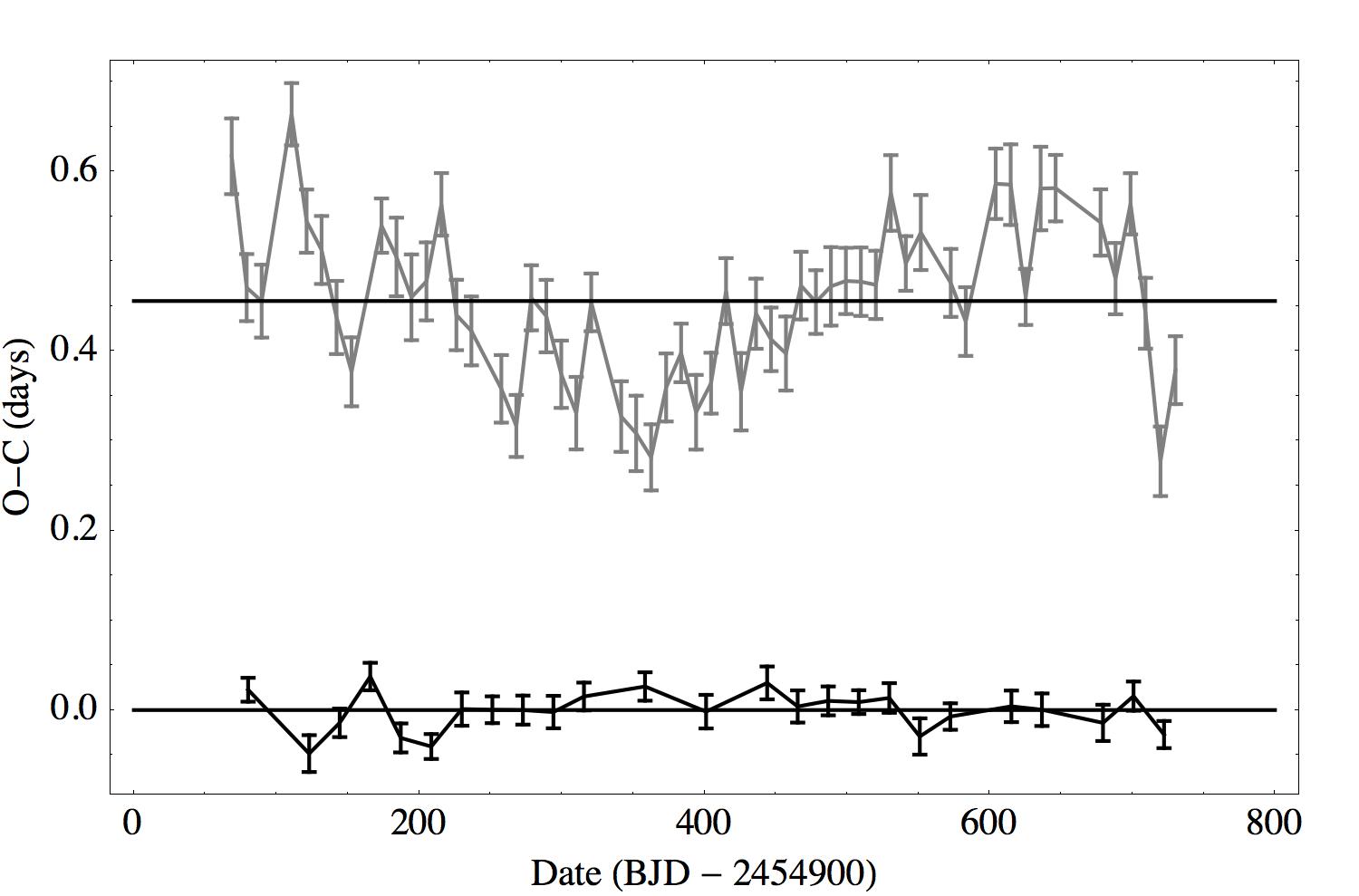}
\caption{Deviations from a constant period for \kepotfo\ (\koiotfo).  The residuals for the outer planet have been displaced vertically for convenience in seeing the TTV signal.}
\label{koi1241}
\end{figure}

\begin{figure}
\includegraphics[width=0.45\textwidth]{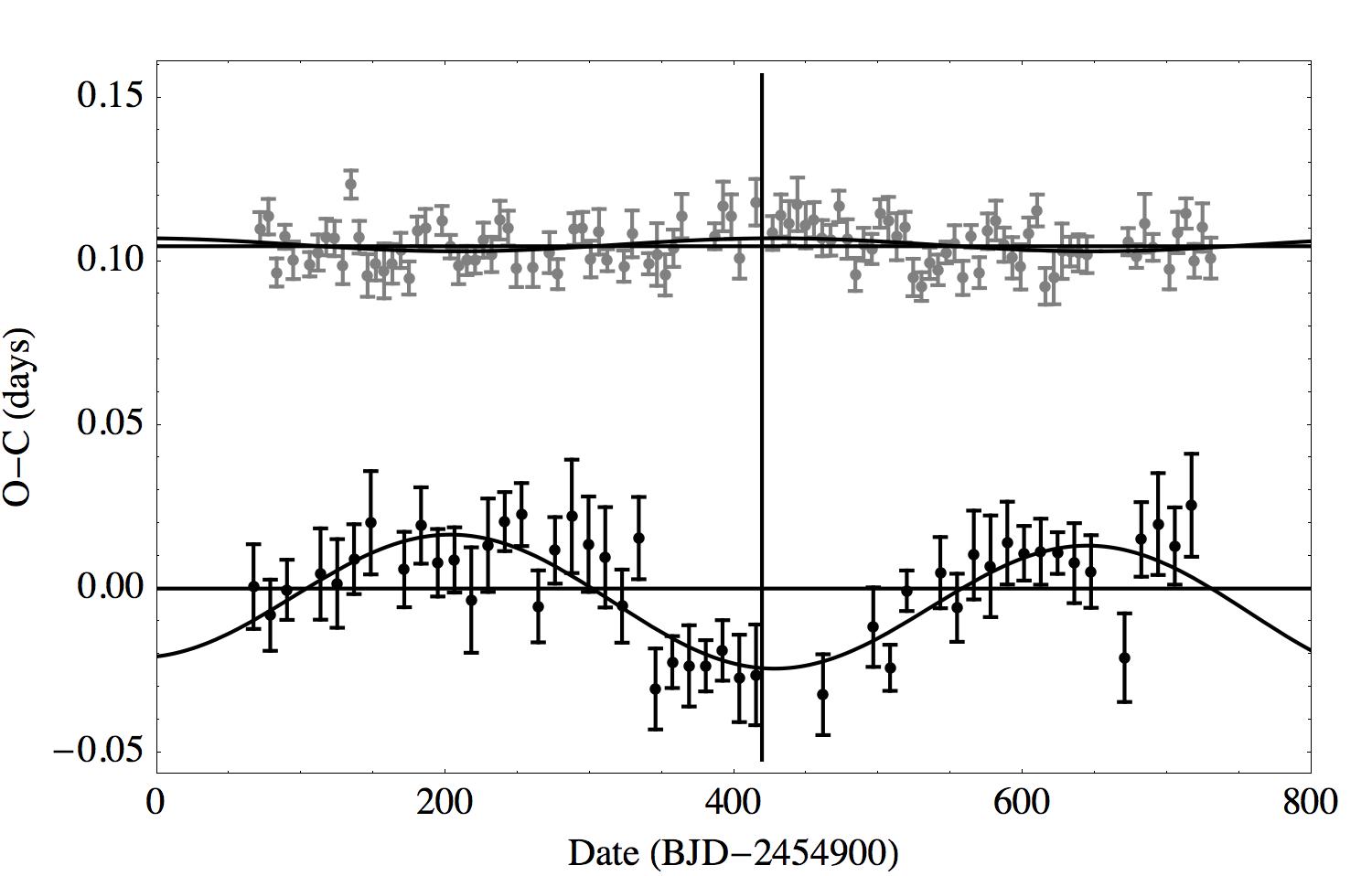}
\caption{Deviations from a constant period for \kepotsz\ (\koiotsz).  The residuals for the outer planet have been displaced vertically for convenience in seeing the TTV signal.  Vertical lines correspond to the times when the line of conjunctions of the two planets crosses the line of sight and are used to measure the TTV phase for the analytic mass estimates shown in Table \ref{tabAnalyticmass}.}
\label{koi1270}
\end{figure}

\begin{figure}
\includegraphics[width=0.45\textwidth]{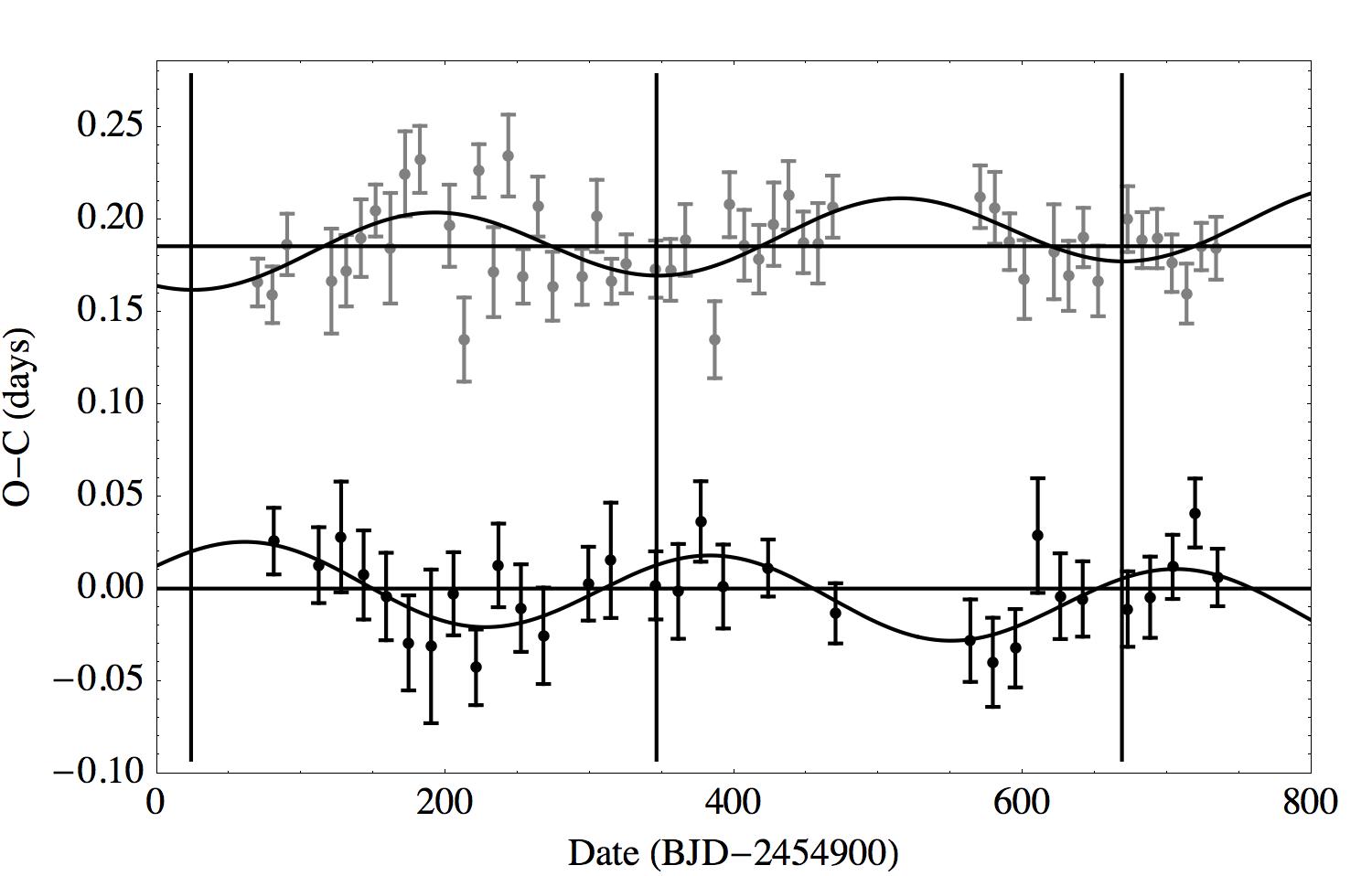}
\caption{Deviations from a constant period for \kepotts\ (\koiotts).  The residuals for the outer planet have been displaced vertically for convenience in seeing the TTV signal.  Vertical lines correspond to the times when the line of conjunctions of the two planets crosses the line of sight and are used to measure the TTV phase for the analytic mass estimates shown in Table \ref{tabAnalyticmass}.}
\label{koi1336}
\end{figure}

\begin{figure}
\includegraphics[width=0.45\textwidth]{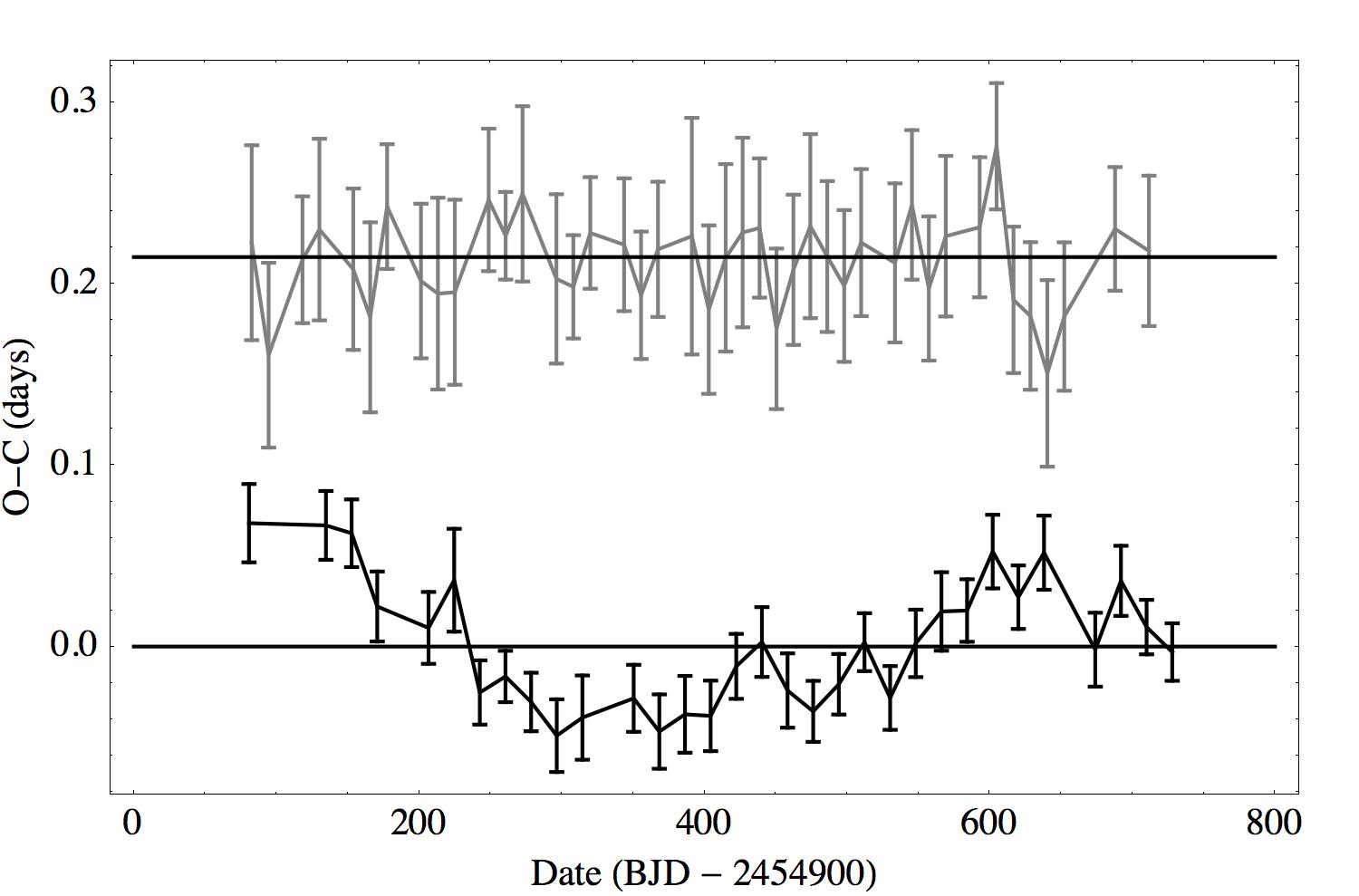}
\caption{Deviations from a constant period for \kepoftn\ (\koioftn).  The residuals for the outer planet have been displaced vertically for convenience in seeing the TTV signal.}
\label{koi1529}
\end{figure}

\begin{figure}
\includegraphics[width=0.45\textwidth]{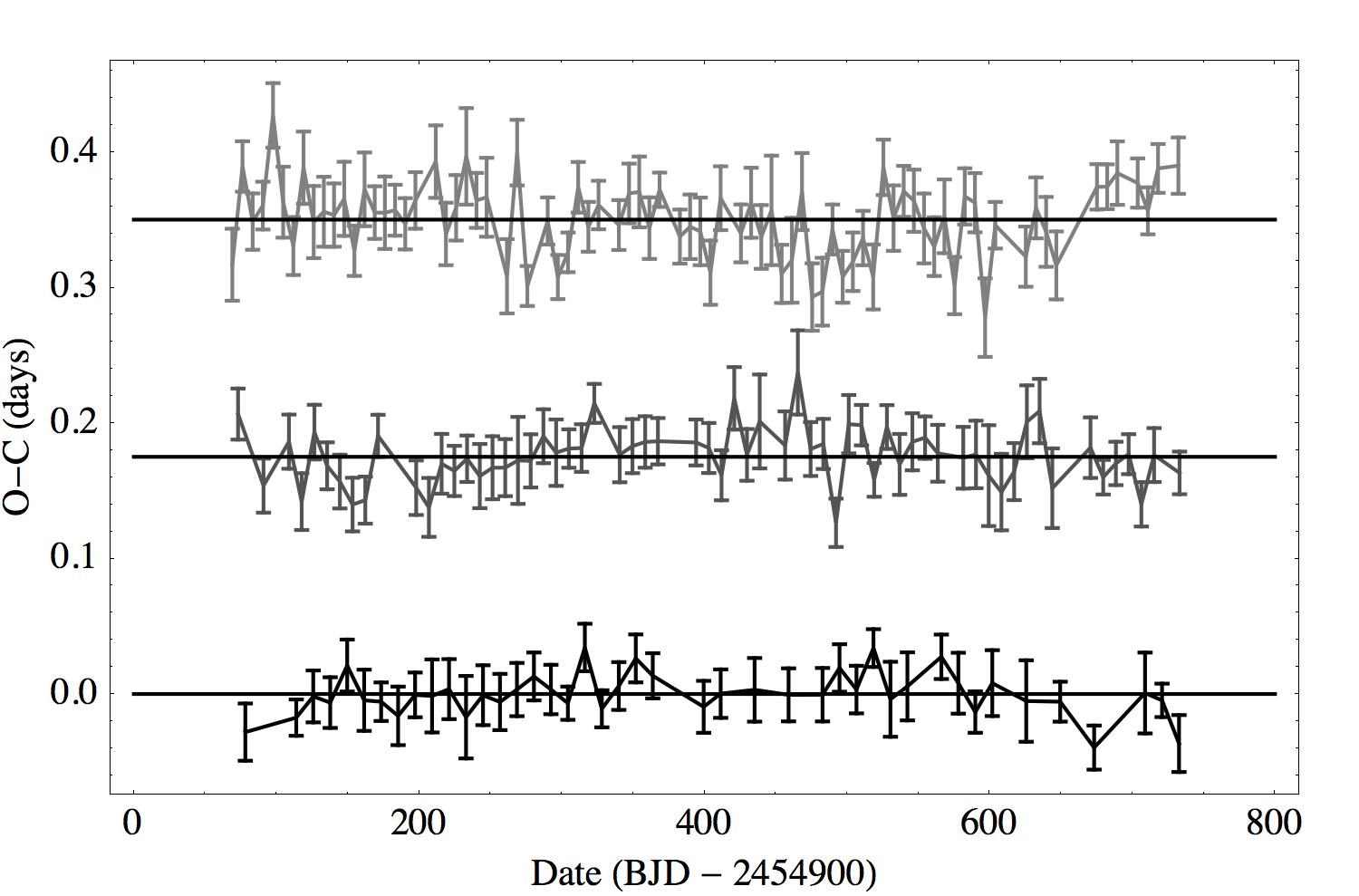}
\caption{Deviations from a constant period for \keptzes\ (\koitzes).  The residuals for the outer planet have been displaced vertically for convenience in seeing the TTV signal.}
\label{koi2086}
\end{figure}

In this appendix we show the data that were used in the analysis of the paper (Figures \ref{koi0148}, \ref{koi0248}, \ref{koi0262}, \ref{koi0620}, \ref{koi0775}, \ref{koi0829}, \ref{koi0886}, \ref{koi0904}, \ref{koi1241}, \ref{koi1270}, \ref{koi1336}, \ref{koi1529}, and \ref{koi2086}).  The timing residuals (TTV signal) of the inner planet(s) are displaced vertically for convenience in viewing.  We also show in Table \ref{tabXimax} the \ximax\ values, the TTV period associated with \ximax , and associated fit parameters (from Equation \ref{fourcomp}) for the pairs of planets that can be confirmed with the Fourier methods.

%Other figures or features, such as ``river plots'' from the photodynamical modelling and TTV phase locations are shown for relevant systems.

%\input{xivalstab}
\begin{table*}
\centering
\begin{minipage}{180mm}
\caption{Results from the \ximax\ calculations for pairs of planets that can be confirmed, or nearly so, with the Fourier method.}
\begin{tabular}{rrrrrrrrr}
\hline
KOI		&	\ximax	&	$P_\text{TTV}$	&	$A$	&	$\sigma_A$	&	$B$	&	$\sigma_B$	&	$C$	&	$\sigma_C$	\\
		&		&	(days)	&	(days)	&	(days)	&	(days)	&	(days)	&	(days)	&	(days)	\\ \hline \hline
148	inner	&	19.6	&	381	&	-0.00202	&	0.00068	&	-0.00191	&	0.00076	&	-0.00032	&	0.00051	\\
	outer	&		&		&	0.00245	&	0.00042	&	0.00044	&	0.00046	&	0.00024	&	0.00031	\\
248	inner	&	69.8	&	386	&	0.00608	&	0.00071	&	0.00102	&	0.00074	&	0.00035	&	0.00052	\\
	outer	&		&		&	-0.01063	&	0.00137	&	-0.00281	&	0.00158	&	0.00063	&	0.00103	\\
262	inner	&	34.8	&	1080	&	0.01328	&	0.00289	&	-0.01321	&	0.00282	&	-0.00909	&	0.00226	\\
	outer	&		&		&	-0.00668	&	0.00188	&	0.00706	&	0.00180	&	0.00482	&	0.00148	\\
620	inner	&	79.6	&	2117	&	-0.02297	&	0.00332	&	-0.00805	&	0.00144	&	0.02013	&	0.00294	\\
	outer	&		&		&	0.09573	&	0.01485	&	0.04132	&	0.00678	&	-0.09856	&	0.01528	\\
775	inner	&	19.1	&	192	&	0.00245	&	0.00150	&	-0.00691	&	0.00150	&	-0.00034	&	0.00106	\\
	outer	&		&		&	-0.00273	&	0.00316	&	0.01003	&	0.00272	&	0.00056	&	0.00207	\\
829	inner	&	10.9	&	527	&	0.00297	&	0.00265	&	-0.00606	&	0.00283	&	-0.00123	&	0.00199	\\
	outer	&		&		&	-0.01049	&	0.00256	&	0.00798	&	0.00269	&	0.00222	&	0.00188	\\
886	inner	&	104.3	&	817	&	0.00676	&	0.00182	&	-0.03673	&	0.00243	&	-0.00956	&	0.00153	\\
	outer	&		&		&	-0.00958	&	0.00526	&	0.04975	&	0.00786	&	0.01469	&	0.00473	\\
904	inner	&	93.2	&	1042	&	-0.04334	&	0.00539	&	0.05861	&	0.00625	&	0.03503	&	0.00461	\\
	outer	&		&		&	0.05047	&	0.00827	&	-0.02519	&	0.00541	&	-0.02724	&	0.00544	\\
1241	inner	&	11.5	&	434	&	0.07482	&	0.01245	&	-0.04064	&	0.01291	&	-0.01055	&	0.00905	\\
	outer	&		&		&	-0.00374	&	0.00586	&	0.01510	&	0.00643	&	0.00415	&	0.00449	\\
1270	inner	&	18.1	&	428	&	0.00000	&	0.00089	&	0.00213	&	0.00103	&	0.00046	&	0.00071	\\
	outer	&		&		&	0.00096	&	0.00198	&	-0.01916	&	0.00218	&	-0.00459	&	0.00152	\\
1336	inner	&	13.6	&	321	&	-0.00904	&	0.00370	&	-0.01294	&	0.00412	&	0.00358	&	0.00283	\\
	outer	&		&		&	0.01880	&	0.00392	&	0.00230	&	0.00450	&	-0.00384	&	0.00300	\\
2086	inner	&	10.0	&	662	&	0.01122	&	0.00412	&	0.00738	&	0.00401	&	0.00006	&	0.00287	\\
	outer	&		&		&	-0.00806	&	0.00335	&	-0.00626	&	0.00342	&	0.00000	&	0.00239	\\ \hline
\end{tabular}
\label{tabXimax}
\end{minipage}
\end{table*}

%\begin{figure}
%\includegraphics[width=0.45\textwidth]{riverkoi088601}
%\caption{River plot for KOI-0886 (\kepees) from the photodynamical model.  The top pair of panels is the time series folded at the mean orbital period with the transit epoch increasing in the vertical direction.  The best fitting model is shown in the middle panels and the residuals following the fit are in the bottom two panels.}
%\label{river0886}
%\end{figure}

%\begin{figure}
%\includegraphics[width=0.45\textwidth]{riverkoi133601}
%\caption{River plot for KOI-1336 (\kepotts) from the photodynamical model.  The top pair of panels is the time series folded at the mean orbital period with the transit epoch increasing in the vertical direction.  The best fitting model is shown in the middle panels and the residuals following the fit are in the bottom two panels.}
%\label{river1336}
%\end{figure}

%\begin{figure}
%\includegraphics[width=0.45\textwidth]{riverkoi152901}
%\caption{River plot for KOI-1529 (\kepoftn) from the photodynamical model.  The top pair of panels is the time series folded at the mean orbital period with the transit epoch increasing in the vertical direction.  The best fitting model is shown in the middle panels and the residuals following the fit are in the bottom two panels.}
%\label{river1529}
%\end{figure}

%\begin{figure}
%\includegraphics[width=0.45\textwidth]{riverkoi208601}
%\caption{River plot for KOI-2086 (\keptzes) from the photodynamical model.  The top pair of panels is the time series folded at the mean orbital period with the transit epoch increasing in the vertical direction.  The best fitting model is shown in the middle panels and the residuals following the fit are in the bottom two panels.}
%\label{river208601}
%\end{figure}

%\begin{figure}
%\includegraphics[width=0.45\textwidth]{riverkoi208602}
%\caption{River plot for KOI-2086 (\keptzes) from the photodynamical model.  The top pair of panels is the time series folded at the mean orbital period with the transit epoch increasing in the vertical direction.  The best fitting model is shown in the middle panels and the residuals following the fit are in the bottom two panels.}
%\label{river208602}
%\end{figure}

%\begin{figure}
%\includegraphics[width=0.45\textwidth]{koi1336masspic}
%\caption{Folded transit light curves for \kepotts\ (\koiotts) centered on the times of transit (TTVs removed).}
%\label{transit1336}
%\end{figure}

\bsp

\label{lastpage}

\end{document}